\documentstyle[11pt,aaspp4]{article}
\lefthead{JENKINS \& PEIMBERT}
\righthead{MOLECULAR HYDROGEN IN THE DIRECTION OF $\zeta$~ORI~A}
\begin{document}
\title{Molecular Hydrogen in the Direction of $\zeta$~Ori~A}
\author{Edward B. Jenkins and Antonio Peimbert}
\affil{Princeton University Observatory\\
Princeton, NJ 08544-1001; ebj@astro.princeton.edu,
antonio@astro.princeton.edu}
\begin{abstract}
A spectrum of $\zeta$~Ori~A over the wavelength interval 950$-$1150\AA\
recorded by Interstellar Medium Absorption Profile Spectrograph (IMAPS)
on the ORFEUS-SPAS~I mission shows Lyman and Werner band absorption
features from molecular hydrogen in rotational levels $J=0$, 1, 2, 3 and
5.  Most of the molecules are found in two distinct velocity components. 
One is at a heliocentric radial velocity of about $-$1~km~s$^{-1}$ with
$\log N({\rm H}_2)=14.5$ and a rotational temperature $T_{\rm
rot}=950$K, while the other is at +25~km~s$^{-1}$ with $\log N({\rm
H}_2)=15.9$ and $T_{\rm rot}=320$K.  Some extra H$_2$ exists in a much
weaker component ($\log N({\rm H}_2)=14.0$) between the two main peaks.  

The H$_2$ component at $-$1~km~s$^{-1}$ exhibits profile shapes that
become broader and show small displacements toward more negative
velocities as $J$ increases.  These changes are inconsistent with a
simple interpretation that uv optical pumping in an optically thin,
uniform medium creates the H$_2$ in excited rotational levels. 
Differential shielding of the uv radiation at certain velocities does
not appear to be a satisfactory explanation for the effect.

Evidence from atomic features at other velocities may offer some insight
on the origin of this unusual behavior exhibited by the H$_2$ profiles.
Absorption features from moderately ionized atoms at $-$94~km~s$^{-1}$
and more highly ionized species at about $-$36~km~s$^{-1}$ suggest that
along the line of sight to $\zeta$~Ori~A there may be a standing bow
shock with an initial compression ratio of 2.6.  This shock is probably
created when a negative-velocity gas flow collides with an obstruction,
in this case a neutral cloud at 0~km~s$^{-1}$.  If this interpretation
is correct, the H$_2$ with the changing profiles may represent molecules
forming in the postshock gas flow that is undergoing further compression
as it recombines and cools.  We suggest that molecules can form
initially by associative detachment of H$^-$ in a moving, warm, partly
ionized medium behind the front.  The H$_2$ in this area is most
conspicuous in the higher $J$ levels.  Later, when the gas becomes very
cool, neutral, and more compressed as it comes nearly to a halt, it is
more easily seen in the lowest $J$ levels.  In this part of the medium,
the principal way of producing H$_2$ should be from reactions on the
surfaces of dust grains, as one expects for quiescent interstellar
clouds.
\end{abstract}

\section{Introduction}\label{intro}

The first detection of hydrogen molecules in space came from a
distinctive pattern of absorption features that appeared in a low
resolution uv spectrum of $\xi$~Per recorded by a spectrometer on a
sounding rocket \markcite{1007} (Carruthers 1970).  Starting with that
pioneering discovery, the Lyman and Werner bands of H$_2$ in the spectra
of early-type stars have led us down a trail of new discoveries about
this most abundant molecule in space.  Progressively more refined
observations by the {\it Copernicus} satellite have given us a
fundamental understanding on this molecule's abundances in various
diffuse cloud environments \markcite{1002, 1176, 1141} (Spitzer et al.
1973; York 1976; Savage et al. 1977), how rapidly it is created and
destroyed in space \markcite{1276} (Jura 1974), and the amount of
rotational excitation that is found in different circumstances
\markcite{1212, 1015, 1772} (Spitzer \& Cochran 1973; Spitzer, Cochran,
\& Hirshfeld 1974; Morton \& Dinerstein 1976).  The observed populations
in excited rotational levels have in turn led to theoretical
interpretations about how this excitation is influenced by such
conditions as the local gas density, temperature and the flux of uv
pumping radiation from nearby stars \markcite{1762, 1925, 1277} (Spitzer
\& Zweibel 1974; Jura 1975a, b).  Many of the highlights of these
investigations have been reviewed by Spitzer \& Jenkins \markcite{1326}
(1975) and Shull \& Beckwith \markcite{2406} (1982).  The Lyman and
Werner bands of H$_2$ can even be used to learn more about the
properties of very distant gas systems whose absorption lines appear in
quasar spectra \markcite{1078, 287} (Foltz, Chaffee, \& Black 1988;
Songaila \& Cowie 1995), although the frequency of finding these H$_2$
features is generally quite low \markcite{2297} (Levshakov et al. 1992).

In addition to the general conclusions just mentioned, there were some
intriguing details that came from the observations of uv absorption
lines.  The early surveys by the {\it Copernicus} satellite indicated
that toward a number of stars the H$_2$ features became broader as the
rotational quantum number $J$ increased \markcite{1212, 1015} (Spitzer
\& Cochran 1973; Spitzer, Cochran, \& Hirshfeld 1974).  An initial
suggestion by Spitzer \& Cochran \markcite{1212} (1973) was that the
extra broadening of the higher $J$ levels could arise from new molecules
that had a large kinetic energy that was liberated as they formed and
left the grain surfaces.  However, a more detailed investigation by
Spitzer \& Morton \markcite{1213} (1976) showed that, as a rule, the
cases that exhibited the line broadening with increasing $J$ were
actually composed of two components that had different rotational
excitations and a velocity separation that was marginally resolved by
the instrument.  In general, they found that the component with a more
negative radial velocity was relatively inconspicuous at low $J$, but
due to its higher rotation temperature it became more important at
higher $J$ and made the composite profile look broader.

By interpreting the rotational populations from the standpoint of
theories on collisional and uv pumping, Spitzer \& Morton
\markcite{1213} (1976) found a consistent pattern where the components
with the most negative velocity in each case had extraordinarily large
local densities and exposure to unusually high uv pumping fluxes.  They
proposed that these components arose from thin, dense sheets of
H$_2$-bearing material in the cold, compressed regions that followed
shock fronts coming toward us.  These fronts supposedly came from either
the supersonic expansions of the stars' H~II regions or perhaps from the
blast waves caused by supernova explosions in the stellar associations.

Now, some twenty years after the original investigations with the {\it
Copernicus} satellite, we have an opportunity to study once again the
behavior of the H$_2$ profiles, but this time with a wavelength
resolution that can cleanly separate the components.  We report here the
results of an investigation of H$_2$ toward $\zeta$~Ori~A, one of the
stars studied earlier that showed the intriguing behavior with the H$_2$
components discussed above.  Once again, the concept of the H$_2$
residing in the dense gas behind a shock comes out as a central theme in
the interpretation, but our description of the configuration given in
\S\ref{shock} is very different from that offered by Spitzer \& Morton
\markcite{1213} (1976).

\section{Observations}\label{obs}

The Lyman and Werner band absorptions of H$_2$ in the spectrum of
$\zeta$~Ori~A were observed with the Interstellar Medium Absorption
Profile Spectrograph (IMAPS).  IMAPS is an objective-grating echelle
spectrograph that was developed in the 1980's as a sounding rocket
instrument \markcite{1390, 1367} (Jenkins et al. 1988, 1989) and was
recently reconfigured to fly in orbit. It can record the spectrum of a
star over the wavelength region 950$-$1150\AA\ at a resolving power of
about 200,000.\footnote{The observations reported here had a resolution
that fell short of this figure, for reasons that are given in
\S\ref{wl_scale}.}  This instrument flew on the ORFEUS-SPAS carrier
launched on 12 September 1993 by the Space Shuttle flight STS-51. 
Jenkins, et al. \markcite{340} (1996) have presented a detailed
description of IMAPS, how it performed during this mission and how the
data were reduced.  Their article is especially useful for pointing out
special problems with the data that were mostly overcome in the
reduction.  It also shows an image of a portion of the echelle spectrum
of $\zeta$~Ori~A.

\begin{figure}
\plotone{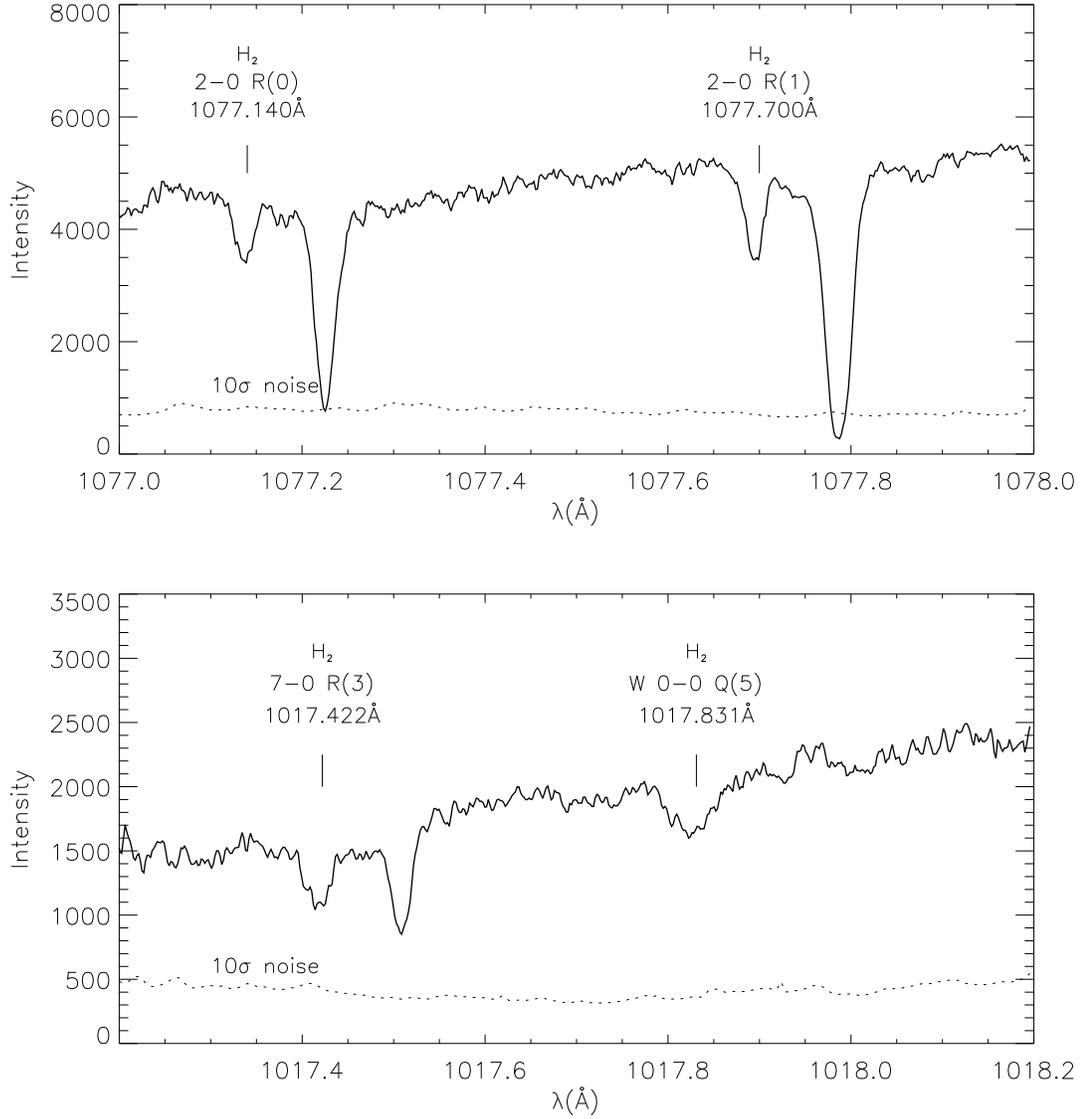}
\caption{Sample wavelength intervals covering four H$_2$
lines in the spectrum of $\zeta$~Ori recorded by IMAPS.  The intensity
scale is in photons per detector pixel.  Except for the transition out
of $J=5$, two prominent velocity components are seen in each H$_2$ line,
one at a heliocentric radial velocity of about $-$1~km~s$^{-1}$ (labeled
with marks that identify the transitions) and another at
+25~km~s$^{-1}$.  Detector pixels are oversampled by a factor of two in
this figure.  Dashed lines show the amplitude of ten times the expected
standard deviation of the points resulting from photon counting
statistics.\label{two_spec}}
\end{figure}

The total exposure time on $\zeta$~Ori was 2412~s, divided among 63
frames, each of which covered \onequarter\ of the echelle's free
spectral range.  Spectra were extracted using an optimal extraction
routine described by Jenkins, et al. \markcite{340} (1996), and
different measurements of the intensity at any given wavelength were
combined with weights proportional to their respective inverse squares
of the errors.  Samples of some very restricted parts of the final
spectrum are shown in Fig.~\ref{two_spec}, where lines from $J$ = 0, 1,
3 and 5 may be seen.

\placefigure{two_spec}

\section{Data Reduction}\label{reduction}

\subsection{Wavelength Scale and Resolution}\label{wl_scale}

Since IMAPS is an objective-grating instrument, there is no way that we
can use an internal line emission light source to provide a calibration
of the wavelength scale.  However, as explained in Jenkins, et al.
\markcite{340} (1996), we have an accurate knowledge of how the apparent
detector coordinates map into real geometrical coordinates on the image
plane, and we also know the focal length of the cross-disperser grating
and the angles of incidence and diffraction for the echelle grating. 
The only unknown parameter that we must measure is a zero offset that is
driven by the pointing of IMAPS relative to the target.  We determined
this offset by measuring the positions of telluric absorption features
of O~I in excited fine-structure levels.  These features are rarely seen
in the interstellar medium, but there is enough atmospheric oxygen above
the orbital altitude of 295~km to produce the absorption features in all
of our spectra.

To obtain a wavelength scale that would give heliocentric
velocities\footnote{To obtain the LSR velocity in the direction of
$\zeta$~Ori~A, one should subtract 17.5~km~s$^{-1}$ from the
heliocentric velocity.  Differential galactic rotation at an assumed
distance of 450~pc to $\zeta$~Ori~A should cause undisturbed gases in
the general vicinity of the star to move at 4.5~km~s$^{-1}$ with respect
to the LSR if the galaxy has a constant rotation velocity of
220~km~s$^{-1}$ and $R_0=8.5$ kpc \protect\markcite{1681} (Gunn, Knapp,
\& Tremaine 1979).  Thus, any feature appearing at a heliocentric
velocity of 22.0~km~s$^{-1}$ should be approximately in the rest frame
of gaseous material in the vicinity of our target.} for all of our
lines, we adjusted the zero offset so that the telluric features
appeared at +27.0~km~s$^{-1}$, a value that was appropriate for the
viewing direction and time of our observations.  The general accuracy of
our wavelength scale is indicated by the fact that oxygen lines in 4
different multiplets all gave velocities within 0.5~km~s$^{-1}$ of the
average.  Also, for H$_2$ lines out of a given $J$ level that had
roughly comparable transition strengths, the dispersion of measured
velocities was about 0.5~km~s$^{-1}$.  The measured position of the
strongest component (for all $J$ levels) of 24.5~km~s$^{-1}$ compares
favorably with the heliocentric velocity of a strong, but complex
absorption feature of Na~I centered on 24~km~s$^{-1}$ \markcite{262}
(Welty, Hobbs, \& Kulkarni 1994).

The excited O~I lines can also be used to give an indication of the
wavelength resolution of our observations.  We measured equivalent
widths of 10.5 and 7.25m\AA\ for the O~I$^*$ and O~I$^{**}$ lines at
1040.9 and 1041.7\AA, respectively.  For the applicable densities and
temperatures of the Earth's upper atmosphere, the occupation of the
singly excited level (O~I$^*$) should be 3 times that of the doubly
excited level (O~I$^{**}$), i.e., their relative numbers are governed by
just their statistical weights $g$.  Making use of this fact allows us
to apply a standard curve of growth analysis to derive
log~N(O~I$^{**}$)~=~14.19 and $b=0.99~{\rm km~s}^{-1}$ (equivalent to a
doppler broadening for $T=950$K).\footnote{These results agree very well
with predictions of the MSIS-86 model of the Earth's thermosphere
\protect\markcite{3267} (Hedin 1987) for the column density and
temperature along a sight line above our orbital altitude and at a
moderate zenith angle (40\arcdeg).} The observed profiles have widths
that correspond to $b=3.0~{\rm km~s}^{-1}$, which leads to the
conclusion that the instrumental spread function is equivalent to a
profile with
\begin{equation}\label{binst}
b_{\rm inst}=\sqrt{3.0^2-1.25^2}=2.7~{\rm km~s}^{-1}
\end{equation}
(The representative $b$ for the excited O~I lines has been elevated to
1.25~km~s$^{-1}$ to account for the small broadening caused by
saturation).

The wavelength resolving power that we obtained is lower than what is
achievable in principle with IMAPS and the pointing stability of the
spacecraft.  We attribute the degradation to small motions of the
echelle grating during the exposures, caused by a sticky bearing that
relieved mechanical stresses at random times.  The magnitude and
character of this effect is discussed in detail by Jenkins, et al.
\markcite{340} (1996).

\subsection{Absorption Line Measurements}\label{line_meas}

We used the MSLAP analysis program\footnote{MSLAP is a third-generation
program developed for NASA.  MSLAP is copyrighted by Charles L. Joseph
and Edward B. Jenkins.} to define the continuum level $I_0$ and
re-express the intensities $I(v)$ in terms of the apparent absorption
optical depths $\tau_a$ as a function of radial velocity,
\begin{equation}\label{tau_a}
\tau_a(v) = \ln \Bigl( {I_0\over I(v)}\Bigr)~.
\end{equation}
For the ideal case where the instrument can resolve the finest details
in velocity, $\tau_a(v)$ usually gives an accurate depiction of a
differential column density per unit velocity through the relation
\begin{equation}\label{N_a}
N_a(v) = 3.768\times 10^{14}{\tau_a(v)\over f\lambda}{\rm cm}^{-2}({\rm
km~s}^{-1})^{-1}~,
\end{equation}
where $f$ is the transition's $f$-value and $\lambda$ is expressed in
\AA.  However, if there are saturated, fine-scale details that are not
resolved, the true optical depths $\tau(v)$ averaged over velocity will
be underestimated, and one will miscalculate the true column density
$N(v)$.  One can ascertain that this is happening if the application of
Eq.~\ref{N_a} for weaker lines indicates the presence of more material
than from the strong ones \markcite{110,3184} (Savage \& Sembach 1991;
Jenkins 1996).  As will be evident in \S\ref{unres_sat}, this appears to
happen for the strongest features of H$_2$ in the $J$ = 0, 1 and 2
levels of rotational excitation.

For $J$ levels 0 through 3, we were able to draw together the results
for many different absorption lines, each going to different rotational
and vibrational levels in the upper electronic states,
2p$\sigma\,B\,^1\Sigma_u^+$ and 2p$\pi\,C\,^1\Pi_u$.  In so doing, it
was important to keep track of the errors in the measured $I(v)$ and
combine redundant information at each velocity in a manner that lowered
the error in the final result.  To achieve this goal, we evaluated for
every individual velocity point the $\chi^2$ from a summation over the
separate transitions,
\begin{equation}\label{chi_sq}
\chi^2(\tau_a)=\sum \biggl( {\exp(-\tau_a) - I/I_0\over
\epsilon(I/I_0)}\biggr)^2~.
\end{equation}

The expected errors in intensity $\epsilon(I/I_0)$ represented a
combination of several sources of error: (1) the noise in the individual
measurements of $I$, (2) an error in the placement of the continuum
$I_0$, and (3) an error in the adopted value of zero spectral intensity
(which is a finite value of real intensity extracted from the echelle
order).  The errors in $I$ (item 1) were measured from the dispersion of
residual intensities on either side of the adopted continuum at points
well removed in velocity from the absorption feature.  This error
generally becomes larger at progressively shorter wavelengths, because
the sensitivity of IMAPS decreases.  (Variations of sensitivity also
result from being away from the center of the echelle blaze function.)
In every case, the noise errors were assumed to be the same magnitude at
low $I/I_0$ at the centers of lines because statistical fluctuations in
the background illumination are important.  (Generally, the background
was about as large as $I_0$, so the noise amplitude would decrease only
by a factor of $\sqrt{2}$.)  In a number of cases, the computed S/N was
higher than 50 (see Tables~\ref{j0table}$-$\ref{j3table}).  Because
there might be some residual systematic errors that we have not
recognized, we felt that it was unwarranted to assume that these cases
had the full reliability as indicated by the calculation of S/N, when
compared with other measurements at lower S/N.  To account for this, we
uniformly adopted an estimate for the relative noise level consistent
with the value
\begin{equation}\label{noise}
{\rm adopted~S/N} = 1/\sqrt{({\rm computed~S/N})^{-2} + 50^{-2}}.
\end{equation}

The error in $I_0$ (item 2 in the above paragraph) represents the
uncertainty of the continuum level that arises from a pure vertical
translation that would be permitted by the noise in the many intensity
measurements that define $I_0$.  It does {\it not} include errors in the
adopted curvature of the continuum [see a discussion of this issue in
the appendix of Sembach \& Savage \markcite{181} (1992)].  For most
cases, the curvature was almost nonexistent. The error in the adopted
background level (item 3) was judged from the dispersion of residual
intensities of saturated atomic lines elsewhere in the spectrum.  At
every velocity point, the worst combinations of the systematic errors
(i.e., both the adopted continuum and background levels are
simultaneously too high or, alternatively, too low) were combined in
quadrature with the random intensity errors (item 1), as modified in
Eq.~\ref{noise}, to arrive at the net $\epsilon(I/I_0)$.

Tables~\ref{j0table}$-$\ref{j3table} show the transitions for the four
lowest rotational levels of H$_2$ covered in our spectrum of
$\zeta$~Ori.  Laboratory wavelengths are taken from the calculated
values of Abgrall, et al. \markcite{280} (1993a) for the Lyman band
system and Abgrall, et al. \markcite{281} (1993b) for the Werner bands. 
Transition $f$-values are from Abgrall \& Roueff \markcite{2066} (1989). 
The listed values of S/N are those computed as described above, but
without the modification from Eq.~\ref{noise}.

All of the lines for $J$ = 4 were too weak to measure.  Only one line
from $J$ = 5 was strong enough to be useful (the Werner 0$-$0\,Q(5) line
at 1017.831\AA\ with $\log (f\lambda)$ = 1.39), although weaker lines
showed very noisy profiles that were consistent with this line.

Many lines (or certain portions thereof) were unsuitable for
measurement.  These lines and the reasons for their rejection are
discussed in the endnotes of the tables.  Table~\ref{j3table} omits some
lines that are far too weak to consider.

\placetable{j0table}
\placetable{j1table}
\placetable{j2table}
\placetable{j3table}
\begin{deluxetable}{
r     % identification
r     % wavelength
r     % log f lambda
r     % s/n
}
\small % this and the following command is to make the endnotes fit!
\tablewidth{400pt}
\tablecaption{Lines from J=0\label{j0table}}
\tablehead{
\colhead{Ident.\tablenotemark{a}} & 
\colhead{$\lambda$ (\AA)} &
\colhead{Log ($f\lambda$)} &
\colhead{S/N} }
\startdata
0$-$0 R(0)\tablenotemark{b}&1108.127&0.275&50\nl
1$-$0 R(0)\phm{\/}&1092.195&0.802&77\nl
2$-$0 R(0)\phm{\/}&1077.140&1.111&46\nl
3$-$0 R(0)\phm{\/}&1062.882&1.282&45\nl
4$-$0 R(0)\phm{\/}&1049.367&1.383&30\nl
5$-$0 R(0)\phm{\/}&1036.545&1.447&39\nl
6$-$0 R(0)\phm{\/}&1024.372&1.473&36\nl
7$-$0 R(0)\phm{\/}&1012.810&1.483&81\nl
8$-$0 R(0)\phm{\/}&1001.821&1.432&32\nl
9$-$0 R(0)\tablenotemark{c}&991.376&1.411&\nodata\nl
10$-$0 R(0)\tablenotemark{d}&981.437&1.314&23\nl
11$-$0 R(0)\tablenotemark{e}&971.985&1.289&\nodata\nl 
12$-$0 R(0)\tablenotemark{d}&962.977&1.098&14\nl
13$-$0 R(0)\tablenotemark{d}&954.412&1.126&20\nl %tablebreak was here
W 0$-$0 R(0)\tablenotemark{f}&1008.552&1.647&31\nl
W 1$-$0 R(0)\tablenotemark{g}&985.631&1.833&\nodata\nl
W 2$-$0 R(0)\tablenotemark{h}&964.981&1.823&\nodata\nl
\enddata
\tablenotetext{a}{All transitions are in the
2p$\sigma\,B\,^1\Sigma_u^+\leftarrow {\rm X}\,^1\Sigma_g^+$ Lyman band
system, unless preceded with a ``W'' which refers to the
2p$\pi\,C\,^1\Pi_u\leftarrow {\rm X}\,^1\Sigma_g^+$ Werner bands.}
\tablenotetext{b}{Not used in the composite profile, because components
1 and 2 were too weak compared with the noise.  For component 3, this
was the weakest line and had the least susceptibility to errors from
saturated substructures.  This line was used to define the preferred
value for $N_{\rm total}$ with Method~A (see \S\protect\ref{method_A}).}
\tablenotetext{c}{Not considered because this line had interference from
the W~1$-$0~P(3) line.}
\tablenotetext{d}{Not included in the composite profile because the S/N
was significantly inferior to those of other lines of comparable log
($f\lambda$).}
\tablenotetext{e}{Stellar flux severely attenuated by the Ly-$\gamma$
feature.}
\tablenotetext{f}{Not included; there is serious interference from the
W~0$-$0~R(1) line.}
\tablenotetext{g}{Not included; there is serious interference from the
W~1$-$0~R(1) line.}
\tablenotetext{h}{Not included; there is serious interference from the
W~2$-$0~R(1) line.}
\end{deluxetable}
\begin{deluxetable}{
r     % identification
r     % wavelength
r     % log f lambda
r     % s/n
}
\small % this and the following command is to make the endnotes fit!
\tablewidth{400pt}
\tablecaption{Lines from J=1\label{j1table}}
\tablehead{
\colhead{Ident.\tablenotemark{a}} & 
\colhead{$\lambda$ (\AA)} &
\colhead{Log ($f\lambda$)} &
\colhead{S/N} }
\startdata
0$-$0 P(1)\tablenotemark{b}&1110.062&$-$0.191&46\nl
1$-$0 P(1)\tablenotemark{c}&1094.052&0.340&40\nl
2$-$0 P(1)\phm{\/}&1078.927&0.624&33\nl
3$-$0 P(1)\phm{\/}&1064.606&0.805&48\nl
4$-$0 P(1)\phm{\/}&1051.033&0.902&48\nl
5$-$0 P(1)\phm{\/}&1038.157&0.956&78\nl
6$-$0 P(1)\tablenotemark{d}&1025.934&0.970&\nodata\nl
7$-$0 P(1)\phm{\/}&1014.325&0.960&62\nl
8$-$0 P(1)\phm{\/}&1003.294&0.931&19\nl
9$-$0 P(1)\tablenotemark{e}&992.808&0.883&14\nl
10$-$0 P(1)\tablenotemark{e}&982.834&0.825&14\nl
11$-$0 P(1)\tablenotemark{e}&973.344&0.759&3\nl
12$-$0 P(1)\tablenotemark{e}&964.310&0.683&12\nl
13$-$0 P(1)\tablenotemark{e}&955.707&0.604&9\nl % tablebreak was here
W 0$-$0 Q(1)\phm{\/}&1009.770&1.384&36\nl
W 1$-$0 Q(1)\tablenotemark{f}&986.796&1.551&8\nl
W 2$-$0 Q(1)\tablenotemark{f}&966.093&1.529&10\nl
0$-$0 R(1)\tablenotemark{g}&1108.632&0.086&39\nl
1$-$0 R(1)\tablenotemark{h}&1092.732&0.618&69\nl
2$-$0 R(1)\phm{\/}&1077.700&0.919&55\nl
3$-$0 R(1)\phm{\/}&1063.460&1.106&59\nl
4$-$0 R(1)\phm{\/}&1049.960&1.225&64\nl
5$-$0 R(1)\phm{\/}&1037.149&1.271&56\nl
6$-$0 R(1)\tablenotemark{i}&1024.986&1.312&11\nl
7$-$0 R(1)\phm{\/}&1013.434&1.307&48\nl
8$-$0 R(1)\phm{\/}&1002.449&1.256&16\nl
9$-$0 R(1)\tablenotemark{j}&992.013&1.252&19\nl
10$-$0 R(1)\tablenotemark{e}&982.072&1.138&14\nl
11$-$0 R(1)\tablenotemark{e}&972.631&1.134&5\nl
12$-$0 R(1)\tablenotemark{e}&963.606&0.829&12\nl
13$-$0 R(1)\tablenotemark{e}&955.064&0.971&9\tablebreak
W 0$-$0 R(1)\tablenotemark{k}&1008.498&1.326&\nodata\nl
W 1$-$0 R(1)\tablenotemark{l}&985.642&1.512&\nodata\nl
W 2$-$0 R(1)\tablenotemark{m}&965.061&1.529&\nodata\nl
\enddata
\tablenotetext{a}{All transitions are in the
2p$\sigma\,B\,^1\Sigma_u^+\leftarrow {\rm X}\,^1\Sigma_g^+$ Lyman band
system, unless preceded with a ``W'' which refers to the
2p$\pi\,C\,^1\Pi_u\leftarrow {\rm X}\,^1\Sigma_g^+$ Werner bands.}
\tablenotetext{b}{For component 3, this was the weakest line and had the
least susceptibility to errors from saturated substructures.  This line
was used to define the preferred value for $N_{\rm total}$.  Component 1
of the 0$-$0~R(2) is near this feature, but it is not close and strong
enough to compromise the measurement of $N_{\rm total}$ with Method~A
(\S\protect\ref{method_A}).  We did not use the line in the composite
profile however.}
\tablenotetext{c}{Not used in the composite profile because of
interference from the 1$-$0~R(2) line.  This interference did not
compromise our use of the line for obtaining a measurement of
Component~3 using Method~B (\S\protect\ref{method_B}).}
\tablenotetext{d}{Stellar flux severely attenuated by the Ly-$\beta$
feature.}
\tablenotetext{e}{Not included in the composite profile because the S/N
was significantly inferior to those of other lines of comparable log
($f\lambda$).}
\tablenotetext{f}{S/N too low to use this line, even though its $\log
(f\lambda)$ is large.}
\tablenotetext{g}{Components 1 and 2 too weak to measure, hence not
included in composite profile.  For Component~3, this line was used in
Method~B (\S\protect\ref{method_B}).}
\tablenotetext{h}{Possible interference from 1092.620 and 1092.990\AA\
lines of S~I, hence not included in composite profile.}
\tablenotetext{i}{On a wing of the stellar Ly-$\beta$, hence the S/N is
low.  Line not used in the composite profile.}
\tablenotetext{j}{Not included in the composite profile because the
error array shows erratic behavior.}
\tablenotetext{k}{This line has interference from the W~0$-$0~R(0) and
8$-$0~P(3) lines.  It was not used.}
\tablenotetext{l}{This line has interference from the W~1$-$0~R(0) line.
It was not used.}
\tablenotetext{m}{This line has interference from the W~2$-$0~R(0) line.
It was not used.}
\end{deluxetable}
\begin{deluxetable}{
r     % identification
r     % wavelength
r     % log f lambda
r     % s/n
}
\small % this and the following command is to make the endnotes fit!
\tablewidth{400pt}
\tablecaption{Lines from J=2\label{j2table}}
\tablehead{
\colhead{Ident.\tablenotemark{a}} & 
\colhead{$\lambda$ (\AA)} &
\colhead{Log ($f\lambda$)} &
\colhead{S/N} }
\startdata
0$-$0 P(2)\tablenotemark{b}&1112.495&$-$0.109&39\nl
1$-$0 P(2)\tablenotemark{c}&1096.438&0.420&51\nl
2$-$0 P(2)\phm{\/}&1081.266&0.706&55\nl
3$-$0 P(2)\tablenotemark{d}&1066.900&0.879&65\nl
4$-$0 P(2)\phm{\/}&1053.284&0.982&35\nl
5$-$0 P(2)\phm{\/}&1040.366&1.017&38\nl
6$-$0 P(2)\tablenotemark{e}&1028.104&1.053&13\nl
7$-$0 P(2)\tablenotemark{f}&1016.458&1.007&34\nl 
8$-$0 P(2)\phm{\/}&1005.390&0.998&29\nl
9$-$0 P(2)\tablenotemark{g}&944.871&0.937&18\nl
10$-$0 P(2)\tablenotemark{h}&984.862&0.907&5\nl
11$-$0 P(2)\tablenotemark{h}&975.344&0.809&7\nl
12$-$0 P(2)\tablenotemark{h}&966.273&0.798&13\nl
13$-$0 P(2)\tablenotemark{h}&957.650&0.662&12\nl % tablebreak was here
W 0$-$0 P(2)\tablenotemark{i}&1012.169&0.746&23\nl
W 1$-$0 P(2)\tablenotemark{h}&989.086&0.904&8\nl
W 2$-$0 P(2)\tablenotemark{h}&968.292&0.843&14\nl
W 0$-$0 Q(2)\phm{\/}&1010.938&1.385&29\nl
W 1$-$0 Q(2)\tablenotemark{j}&987.972&1.551&7\nl
W 2$-$0 Q(2)\tablenotemark{j}&967.279&1.530&11\nl
0$-$0 R(2)\tablenotemark{k}&1110.119&0.018&45\nl
1$-$0 R(2)\tablenotemark{l}&1094.243&0.558&56\nl
2$-$0 R(2)\tablenotemark{c}&1079.226&0.866&35\nl
3$-$0 R(2)\phm{\/}&1064.994&1.069&53\nl
4$-$0 R(2)\phm{\/}&1051.498&1.168&76\nl
5$-$0 R(2)\phm{\/}&1038.689&1.221&72\nl
6$-$0 R(2)\tablenotemark{m}&1026.526&1.267&\nodata\nl
7$-$0 R(2)\phm{\/}&1014.974&1.285&52\nl
8$-$0 R(2)\phm{\/}&1003.982&1.232&40\nl
9$-$0 R(2)\phm{\/}&993.547&1.228&20\nl
10$-$0 R(2)\phm{\/}&983.589&1.072&18\nl
11$-$0 R(2)\tablenotemark{h}&974.156&1.103&4\nl
12$-$0 R(2)\tablenotemark{n}&965.044&0.161&\nodata\nl
13$-$0 R(2)\tablenotemark{h}&956.577&0.940&10\nl
W 0$-$0 R(2)\phm{\/}&1009.024&1.208&32\nl
W 1$-$0 R(2)\tablenotemark{h}&986.241&1.409&3\tablebreak
W 2$-$0 R(2)\phm{\/}&965.791&1.490&16\nl
\enddata
\tablenotetext{a}{All transitions are in the
2p$\sigma\,B\,^1\Sigma_u^+\leftarrow {\rm X}\,^1\Sigma_g^+$ Lyman band
system, unless preceded with a ``W'' which refers to the
2p$\pi\,C\,^1\Pi_u\leftarrow {\rm X}\,^1\Sigma_g^+$ Werner bands.}
\tablenotetext{b}{Component 1 of this line is too weak to see above the
noise, and Component 3 has interference from Component 1 of the
0$-$0~R(3) line.  Hence this transition is not useful.}
\tablenotetext{c}{In constructing the composite profile, we used only
the velocity interval covering Component 3 because Components 1 and 2
are completely buried in the noise.}
\tablenotetext{d}{Component 1 feature seems to be absent for some reason
that is not understood.  Perhaps an unidentified feature on the edge of
this component makes it unrecognizable.}
\tablenotetext{e}{Stellar flux severely attenuated by the Ly-$\beta$
feature.  This line was not used because the S/N was too low.}
\tablenotetext{f}{Component 1 was badly corrupted by an unidentified
line. Only the region around Component~3 was used.}
\tablenotetext{g}{The nearby W~1$-$0~Q(5) line makes the continuum
uncertain.  Thus, we did not use the 9$-$0 P(2) line.}
\tablenotetext{h}{Not included in the composite profile because the S/N
was significantly inferior to those of other lines of comparable log
($f\lambda$).}
\tablenotetext{i}{This line was not used because it might be corrupted
by the presence of the 1012.502\AA\ line of S~III at $-$80~km~s$^{-1}$.}
\tablenotetext{j}{S/N too low to use this line, even though its $\log
(f\lambda)$ is large.}
\tablenotetext{k}{We used only the portion covered by Component 3, since
Component 1 of this line has serious interference from Component 3 of
the 0$-$0~P(1) line.}
\tablenotetext{l}{We used only the portion covered by Component 3, since
the continuum just to the left of Component 1 is compromised by the
presence of Component 3 of the 1$-$0~P(1) line.}
\tablenotetext{m}{Stellar flux severely attenuated by the Ly-$\beta$
feature.  This line was not used.}
\tablenotetext{n}{There is interference from the N~I line at 965.041\AA. 
Hence this line was not used.}
\end{deluxetable}
\begin{deluxetable}{
r     % identification
r     % wavelength
r     % log f lambda
r     % s/n
}
\small % this and the following command is to make the endnotes fit!
\tablewidth{400pt}
\tablecaption{Lines from J=3\label{j3table}}
\tablehead{
\colhead{Ident.\tablenotemark{a}} & 
\colhead{$\lambda$ (\AA)} &
\colhead{Log ($f\lambda$)} &
\colhead{S/N} }
\startdata
0$-$0 P(3)\tablenotemark{b}&1115.895&$-$0.083&45\nl
1$-$0 P(3)\tablenotemark{b}&1099.787&0.439&31\nl
2$-$0 P(3)\tablenotemark{c}&1084.561&0.734&\nodata\nl
3$-$0 P(3)\phm{\/}&1070.141&0.910&26\nl
4$-$0 P(3)\phm{\/}&1056.472&1.006&56\nl
5$-$0 P(3)\phm{\/}&1043.502&1.060&48\nl
6$-$0 P(3)\phm{\/}&1031.192&1.055&41\nl
7$-$0 P(3)\phm{\/}&1019.500&1.050&57\nl 
8$-$0 P(3)\tablenotemark{d}&1008.383&1.004&\nodata\nl 
9$-$0 P(3)\tablenotemark{e}&997.824&0.944&18\nl
10$-$0 P(3)\tablenotemark{e}&987.768&0.944&10\nl
11$-$0 P(3)\tablenotemark{e}&978.217&0.817&20\nl
12$-$0 P(3)\tablenotemark{e,f}&969.089&0.895&10\nl
13$-$0 P(3)\tablenotemark{e}&960.449&0.673&12\nl % tablebreak was here
W 0$-$0 P(3)\phm{\/}&1014.504&0.920&54\nl
W 1$-$0 P(3)\tablenotemark{g}&991.378&1.075&\nodata\nl
W 2$-$0 P(3)\tablenotemark{e}&970.560&0.974&10\nl
W 0$-$0 Q(3)\phm{\/}&1012.680&1.386&31\nl
W 1$-$0 Q(3)\tablenotemark{h}&989.728&1.564&\nodata\nl
W 2$-$0 Q(3)\tablenotemark{i,j}&969.047&1.530&8\nl
0$-$0 R(3)\tablenotemark{k}&1112.582&$-$0.024&\nodata\nl
1$-$0 R(3)\tablenotemark{l}&1096.725&0.531&\nodata\nl
2$-$0 R(3)\phm{\/}&1081.712&0.840&47\nl
3$-$0 R(3)\phm{\/}&1067.479&1.028&42\nl
4$-$0 R(3)\phm{\/}&1053.976&1.137&37\nl
5$-$0 R(3)\phm{\/}&1041.157&1.222&49\nl
6$-$0 R(3)\phm{\/}&1028.985&1.243&24\nl
7$-$0 R(3)\phm{\/}&1017.422&1.263&35\nl
8$-$0 R(3)\phm{\/}&1006.411&1.207&18\nl
9$-$0 R(3)\phm{\/}&995.970&1.229&33\nl
10$-$0 R(3)\tablenotemark{e}&985.962&0.908&4\nl
11$-$0 R(3)\tablenotemark{m}&976.551&1.104&\nodata\nl
12$-$0 R(3)\tablenotemark{e}&967.673&1.347&10\nl
13$-$0 R(3)\tablenotemark{e}&958.945&0.931&10\nl
W 0$-$0 R(3)\tablenotemark{n}&1010.129&1.151&40\nl
W 1$-$0 R(3)\tablenotemark{i}&987.445&1.409&6\tablebreak
W 2$-$0 R(3)\tablenotemark{e}&966.780&0.883&16\nl
\enddata
\tablenotetext{a}{All transitions are in the
2p$\sigma\,B\,^1\Sigma_u^+\leftarrow {\rm X}\,^1\Sigma_g^+$ Lyman band
system, unless preceded with a ``W'' which refers to the
2p$\pi\,C\,^1\Pi_u\leftarrow {\rm X}\,^1\Sigma_g^+$ Werner bands.}
\tablenotetext{b}{This line is too weak to show up above the noise.  It
was not used in constructing the composite profile.}
\tablenotetext{c}{This line could not be used because it has serious
interference from the 1084.562 and 1084.580\AA\ lines from an excited
fine-structure level of N~II.}
\tablenotetext{d}{Not used since this line has interference from the
W~0$-$0\,R(1) line.}
\tablenotetext{e}{Not included in the composite profile because the S/N
was significantly inferior to those of other lines of comparable log
($f\lambda$).}
\tablenotetext{f}{Not used since this line has interference from the
W~2$-$0\,Q(3) line.}
\tablenotetext{g}{Not used since this line has interference from the
9$-$0\,R(0) line.}
\tablenotetext{h}{Line is submerged in a deep stellar line of N~III at
989.8\AA.  Thus, it could not be used.}
\tablenotetext{i}{S/N too low to use this line, even though its log
($f\lambda$) is large.}
\tablenotetext{j}{Not used since this line has interference from the
12$-$0\,P(3) line.}
\tablenotetext{k}{This line could not be used because it has
interference from the 0$-$0\,P(2) line.}
\tablenotetext{l}{This line could not be used because it has
interference from the 1096.877\AA\ line of Fe~II.}
\tablenotetext{m}{The left-hand side of Component~1 has interference
from Component~3 of the line of O~I at 976.448\AA.  This line could not
be used even for Component~3 because the continuum level was uncertain.}
\tablenotetext{n}{Line inadvertently omitted.  The omission was
discovered long after the combined analysis had been completed.}
\end{deluxetable}
\clearpage

\section{Results}\label{results}

\begin{figure}
\plotone{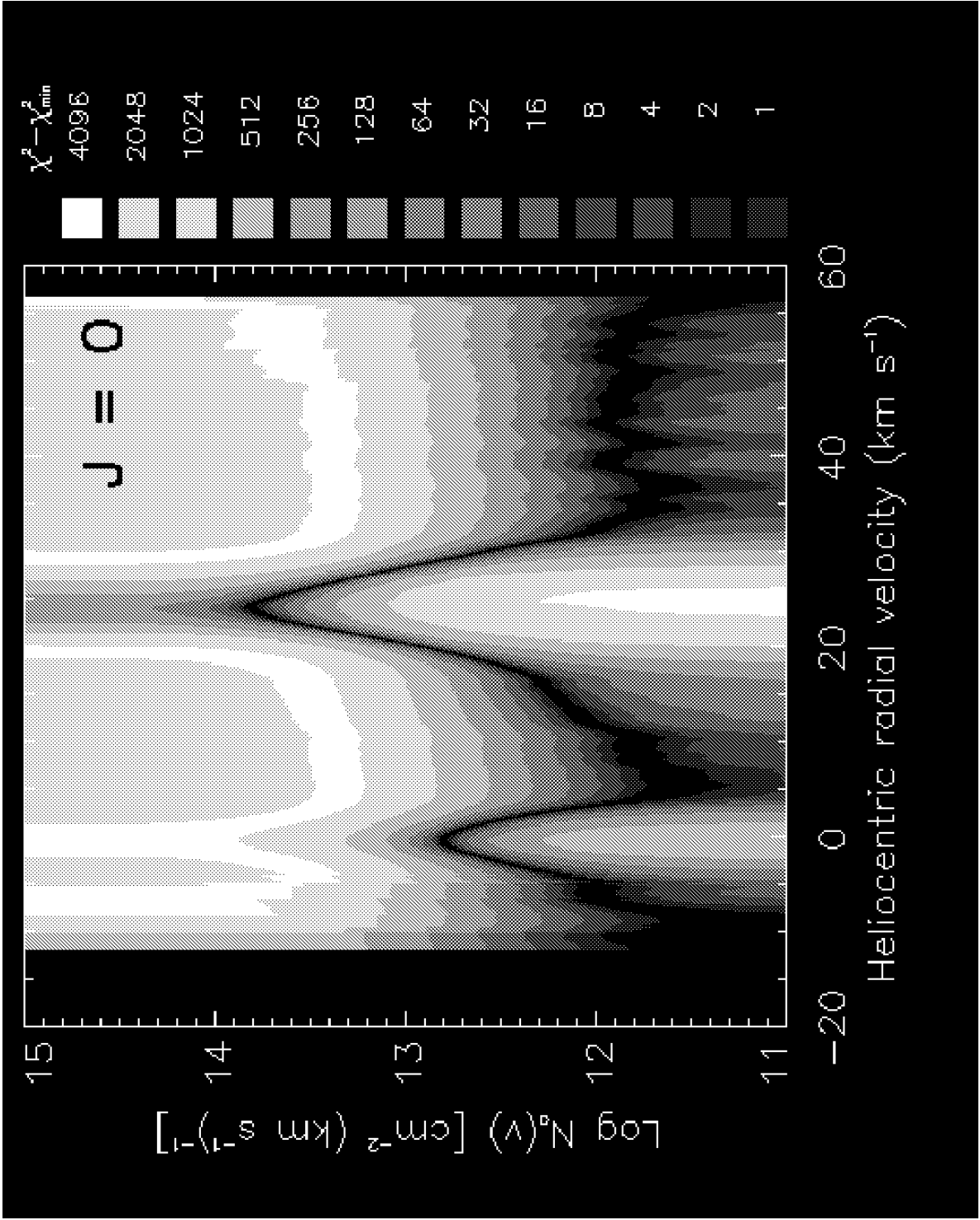}
\caption{A composite of 8 absorption profiles from H$_2$ in
the $J$ = 0 rotational level.  Transitions listed in
Table~\protect\ref{j0table} were used, except where noted.  Shades of
gray, as indicated by the boxes on the right, map out the changes in
$\chi^2-\chi_{\rm min}^2$ as a function of $\log N_a(v)$ for each value
of $v$.  For reasons discussed in \S\protect\ref{unres_sat} the strong
peak on the right-hand side probably understates the true amount of
H$_2$ that is really present.\label{j0fig}}
\end{figure}

\begin{figure}
\plotone{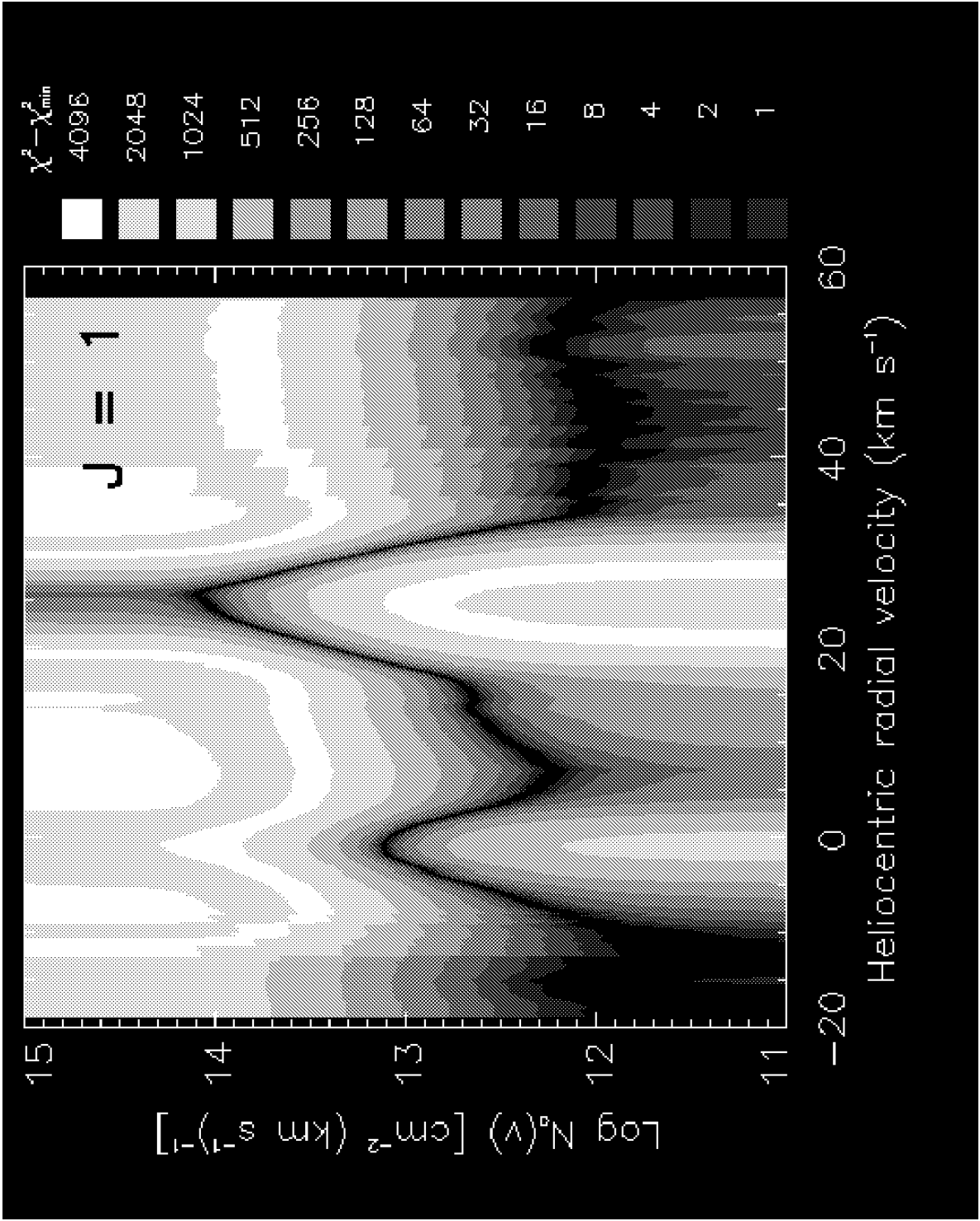}
\caption{Same as for Fig.~\protect\ref{j0fig}, except that
the applicable transitions, listed in Table~\protect\ref{j1table}, are
from the $J$ = 1 level.  Thirteen transitions were combined to make this
figure.  As with Fig.~\protect\ref{j0fig}, the rightmost, strong peak
probably under-represents the true amount of H$_2$.\label{j1fig}}
\end{figure}

\begin{figure}
\plotone{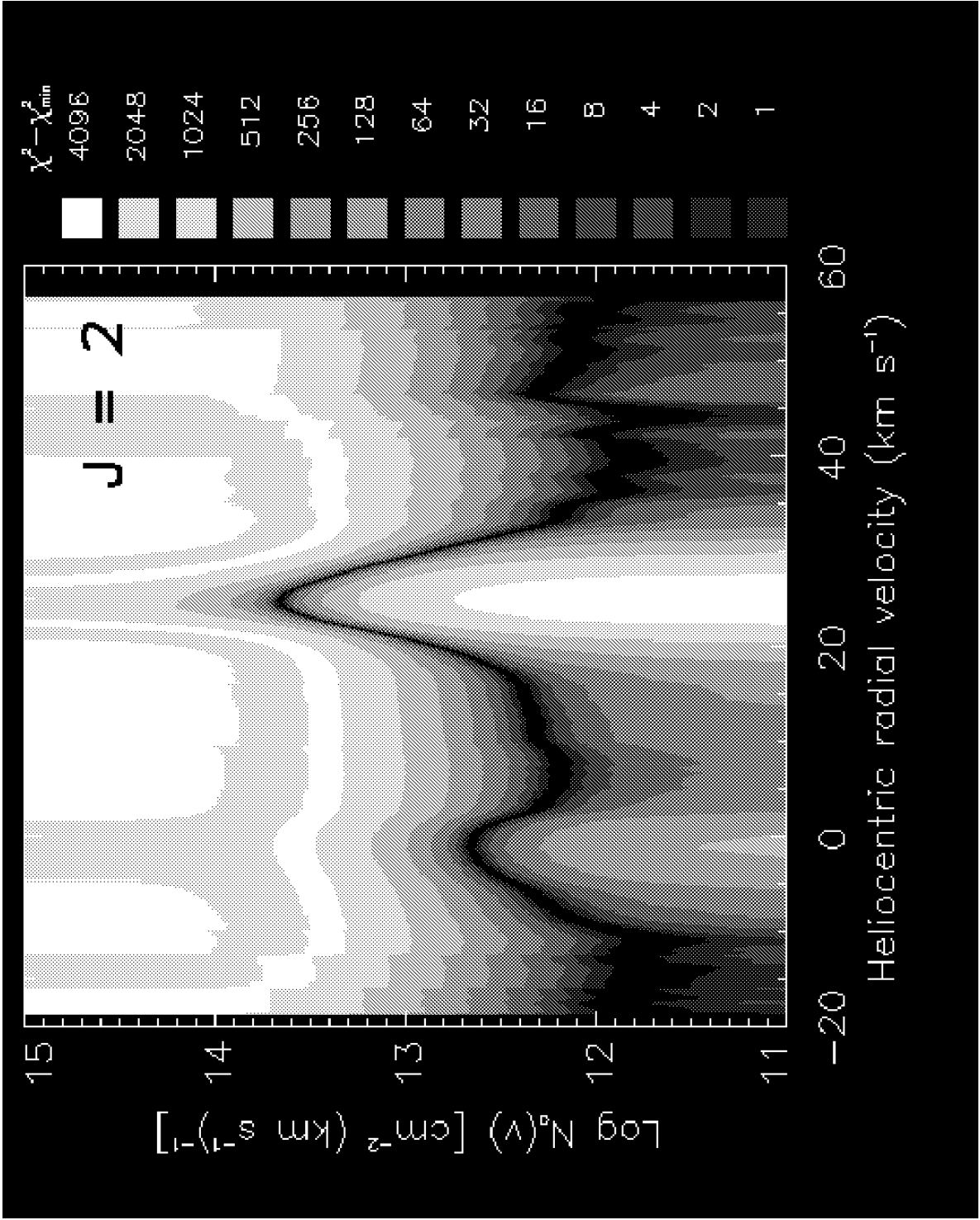}
\caption{Same as for Fig.~\protect\ref{j0fig}, except that
the applicable transitions, listed in Table~\protect\ref{j2table}, are
from the $J$ = 2 level.  A total of 19 transitions were used to
construct this figure, but because of interference problems only 14 of
them covered Components 1 and 2.\label{j2fig}}
\end{figure}

\begin{figure}
\plotone{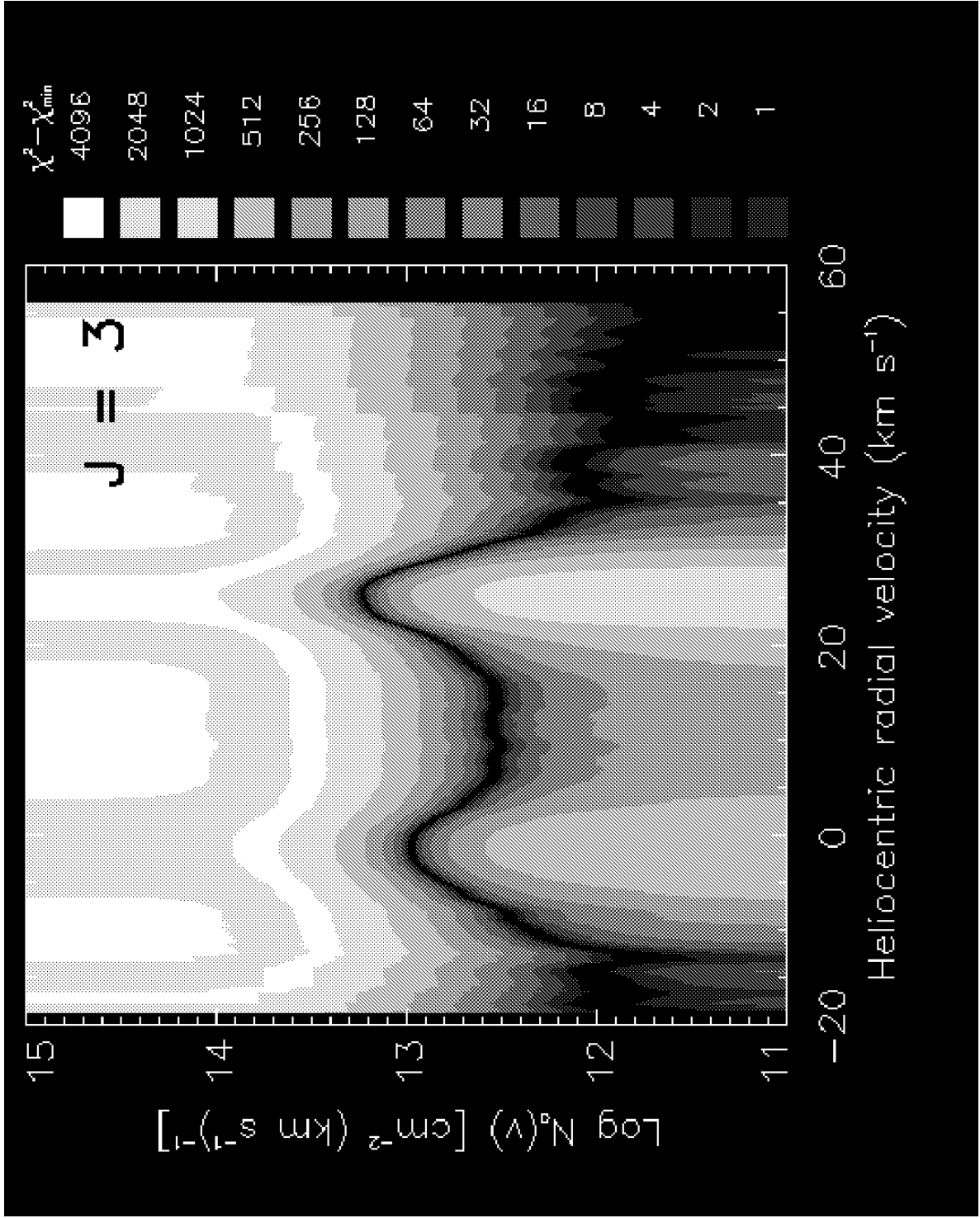}
\caption{Same as for Fig.~\protect\ref{j0fig}, except that
the 15 applicable transitions, listed in Table~\protect\ref{j3table},
are from the $J$ = 3 level.  Unlike the cases for $J$ = 0, 1 or 2, the
right-hand peak does not show any disparities in the height from one
transition to another, indicating that the representation is probably
correct.\label{j3fig}}
\end{figure}

\begin{figure}
\plotone{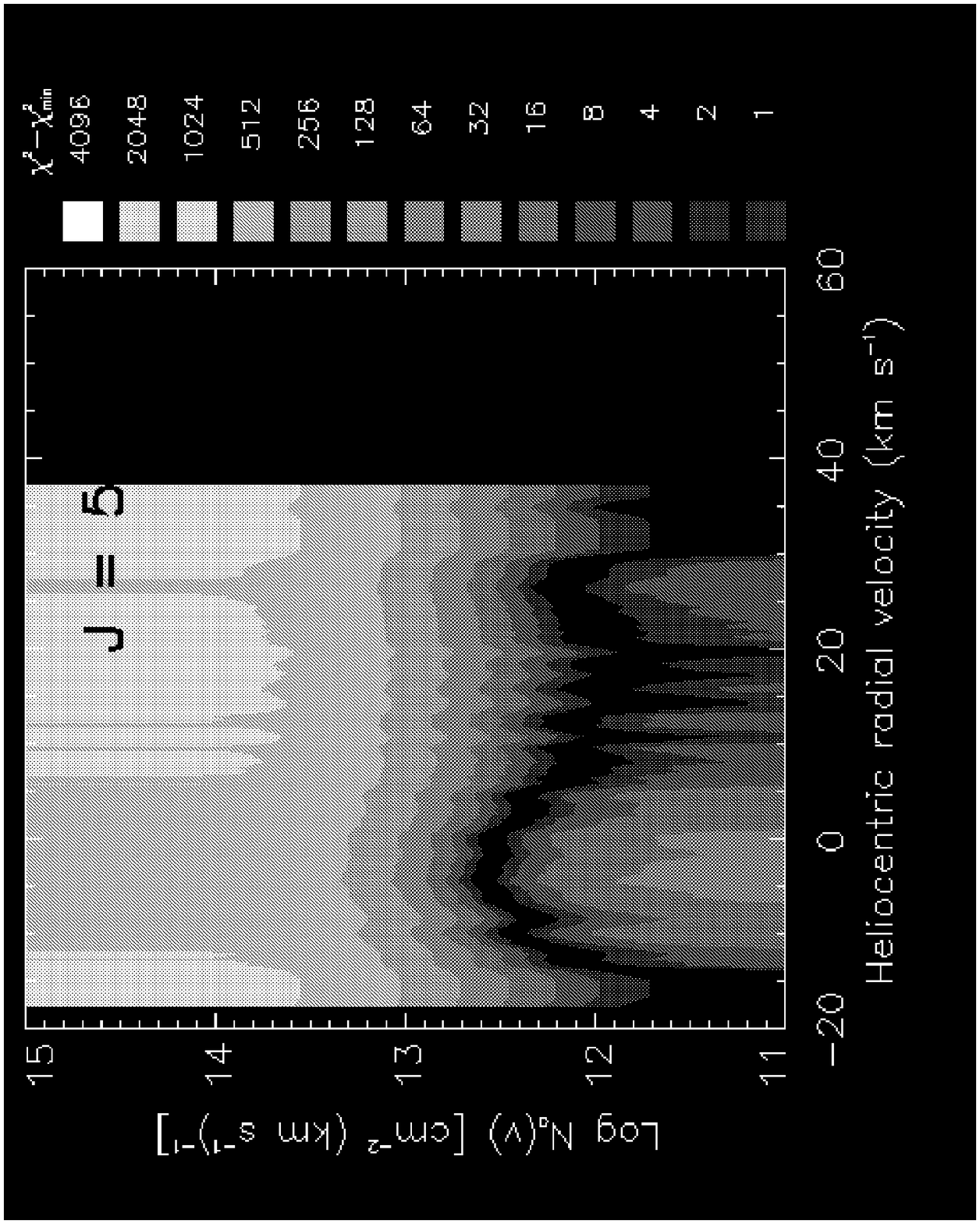}
\caption{Same as for Fig.~\protect\ref{j0fig}, except that
only one transition, the Werner 0$-$0~Q(5) line, was used.
\label{j5fig}}
\end{figure}

Figs.~\ref{j0fig}$-$\ref{j5fig} show gray-scale representations of
$\chi^2-\chi_{\rm min}^2$ as a function of $\log N_a(v)$ and the
heliocentric radial velocity $v$.  The minimum value $\chi_{\rm min}^2$
was determined at each velocity, and our representation that shows how
rapidly $\chi^2$ increases on either side of the most probable $\log
N_a(v)$ (i.e., the value where $\chi_{\rm min}^2$ is achieved) is a
valid measure of the relative confidence of the result \markcite{1666}
(Lampton, Margon, \& Bowyer 1976).  Since we are measuring a single
parameter, the $\chi^2$ distribution function with 1 degree of freedom
is appropriate, and thus, for example, 95\% of the time we expect the
true intensity to fall within a band where $\chi^2-\chi_{\rm min}^2 <
3.8$, i.e., the ``$\pm 2\sigma$'' zone.  To improve on the range of the
display without sacrificing detail for low values of $\chi^2-\chi_{\rm
min}^2$, the actual darknesses in the figures and their matching
calibration squares on the right are scaled to the quantity
$\log(1+\chi^2-\chi_{\rm min}^2)$. Measurements at velocities separated
by more than a single detector pixel (equivalent to 1.25 km~s$^{-1}$)
should be statistically independent.\footnote{This statement is not
strictly true, since single photoevents that fall near the border of two
pixels will contribute a signal to each one.  However, the width of one
pixel is a reasonable gauge for distance between nearly independent
measurements if one wants to judge the significance of the $\chi^2$'s.} 
This separation is less than the wavelength resolving power however. 
Thus, reasonable assumptions about the required continuity of the
profiles for adjacent velocities can, in principle, restrict the range
of allowable departures from the minimum $\chi^2$ even further than the
formal confidence limits.

\placefigure{j0fig}
\placefigure{j1fig}
\placefigure{j2fig}
\placefigure{j3fig}
\placefigure{j5fig}

The profiles that appear in Figs.~\ref{j0fig}$-$\ref{j5fig} indicate
that there are two prominent peaks in H$_2$ absorption, with the
left-hand one holding molecules with a higher rotational temperature
than the one on the right.  This effect, one that creates dramatic
differences in the relative sizes of the two peaks with changing $J$,
was noted earlier by Spitzer,  et al. \markcite{1015} (1974).  There is
also some H$_2$ that spans the velocities between these two peaks.  For
the purposes of making some general statements about the H$_2$, we
identify the material that falls in the ranges $-$15 to +5, +5 to +15,
and +15 to +35 km~s$^{-1}$ as Components 1, 2 and 3, respectively. 
While some residual absorption seems to appear outside the ranges of the
3 components, we are not sure of its reality.  Some transitions seemed
to show convincing extra absorption at these large velocities, while
others did not.

Component~1 shows a clear broadening as the profiles progress from $J=0$
to 5. Precise determinations of this effect and the accompanying
uncertainties in measurement will be presented in \S\ref{prof_changes}. 
The widths of the profiles for Component~3 also seem to increase with
$J$, but the effect is not as dramatic as that shown for Component~1. 
We are reluctant to present any formal analysis of the broadening for
Component~3 because we believe the $N_a(v)$ profile shapes misrepresent
the true distributions of molecules with velocity for $J=0$, 1 and 2,
for reasons given in \S\ref{unres_sat}.  As a rough indication of the
trend, we state here only that the {\it apparent} profile widths are
4.5, 5.8, 5.8 and $7.7~{\rm km~s}^{-1}$ (FWHM) for $J=0$, 1, 2 and 3,
respectively.  These results for this component only partly agree with
the finding by Spitzer,  et al. \markcite{1015} (1974) that the velocity
width of molecules in the $J=1$ state is higher than those in both $J=0$
or $J=2$.  The latter conclusions were based on differences in the $b$
parameters of the curves of growth for the lines.

Table~\ref{comp_summary} lists our values for the column densities $\int
N_a(v)\,dv$, obtained for profiles that follow the valley of minimum
$\chi^2$.  Exceptions to this way of measuring $N({\rm H}_2)$ are
discussed in \S\ref{unres_sat} below.  We also list in the table the
results that were obtained by Spitzer,  et al. \markcite{1015} (1974)
and Spitzer \& Morton \markcite{1213} (1976).  With only two significant
exceptions, our results seem to be in satisfactory agreement with these
previous determinations.  One of the discrepancies is the difference
between our determination $\log N(J=5)=13.70$ for Component~1, compared
with the value of 13.32 found by Spitzer, et al. \markcite{1015} (1974). 
We note that latter was based on lines that had special problems: either
the lines had discrepant velocities or the components could not be
resolved.  The second discrepancy is between our value of $\log
N(J=0)=15.09$ (Method~A discussed in \S\ref{method_A}) or 14.79
(Method~B given in \S\ref{method_B}) for Component~3 and the value 15.77
found by Spitzer \& Morton \markcite{1213} (1976) from an observation of
just the Lyman 4$-$0\,R(0) line.  However, this line is very badly
saturated (the central optical depth must be about 12 with {\it our}
value of $\log N$ and $b=2~{\rm km~s}^{-1}$), and thus it is not
suitable, by itself, for measuring a column density.

\placetable{comp_summary}
\begin{deluxetable}{
c    % J
c    % comp 1
c    % comp 2
c    % comp 3
}
\small
\tablewidth{0pt}
\tablecaption{Log Column Densities\tablenotemark{a}~~and Rotational
Temperatures\label{comp_summary}}
\tablehead{
\colhead{} & 
\colhead{Component 1} &
\colhead{Component 2\tablenotemark{b}} &
\colhead{Component 3} \\
\colhead{$J$} &
\colhead{($-15 < v < +5$ km~s$^{-1}$)} &
\colhead{($+5 < v < +15$ km~s$^{-1}$)} &
\colhead{($+15 < v < +35$ km~s$^{-1}$)} }
\startdata
0&13.53 (13.46B\tablenotemark{c}, 13.48\tablenotemark{d}~)&12.86
(13.23\tablenotemark{d}~)&15.09\tablenotemark{e}, 14.79\tablenotemark{f}
(15.21A\tablenotemark{c}, 15.77\tablenotemark{d}~)\nl
1&13.96 (14.15B\tablenotemark{c}, 14.20\tablenotemark{d}~)&13.44
(13.85\tablenotemark{d}~)&15.72\tablenotemark{e}, 15.69\tablenotemark{f}
(15.43B\tablenotemark{c}~)\nl
2&13.64 (13.64B\tablenotemark{c}, 13.68\tablenotemark{d}~)&13.27
(13.36\tablenotemark{d}~)&14.78\tablenotemark{e}, 14.66\tablenotemark{f}
(14.74B\tablenotemark{c}, 14.87\tablenotemark{d}~)\nl
3&13.99 (14.05A\tablenotemark{c}, 14.08\tablenotemark{d}~)&13.55
(13.69\tablenotemark{d}~)&14.19 (14.14A\tablenotemark{c},
14.34\tablenotemark{d}~)\nl
4&\nodata (13.22A\tablenotemark{c}, 13.11\tablenotemark{d}~)&\nodata
($-\infty$\tablenotemark{d}~)&\nodata (12.95A\tablenotemark{c},
12.85\tablenotemark{d}~)\nl
5&13.70 (13.32A\tablenotemark{c},
13.45\tablenotemark{d}~)&13.13\tablenotemark{g}
(12.48\tablenotemark{d}~)&13.21
($<$12.79\tablenotemark{c},$-\infty$\tablenotemark{d}~)\nl
Total&14.52&14.01&15.86\tablenotemark{e}, 15.79\tablenotemark{f}\nl
Rot. Temp.\tablenotemark{h}&950K&960K&320K\tablenotemark{e},
340K\tablenotemark{f}\nl
\enddata
\tablenotetext{a}{Numbers in parentheses are from earlier {\it
Copernicus} results reported by Spitzer, Cochran \& Hirshfeld
\markcite{1015} (1974) and Spitzer \& Morton \markcite{1213} (1976) for
comparison with our results (and to fill in for $J$ = 4).}
\tablenotetext{b}{Not really a distinct component, but rather material
that seems to bridge the gap between Components 1 and 3.}
\tablenotetext{c}{From Spitzer, Cochran \& Hirshfeld \markcite{1015}
(1974), with errors A = 0.04$-$0.09 and B = 0.10$-$0.19.}
\tablenotetext{d}{From  Spitzer \& Morton \markcite{1213} (1976).}
\tablenotetext{e}{Derived from Method A discussed in
\S\protect\ref{method_A}.}
\tablenotetext{f}{Derived from Method B discussed in
\S\protect\ref{method_B}.}
\tablenotetext{g}{Not a distinct component (see
Fig.~\protect\ref{j5fig}).  The number given is a formal integration
over the specified velocity range and represents the right-hand wing of
the very broad component centered near the velocity of Component~1.}
\tablenotetext{h}{From the reciprocal of the slope of the best fit to
$\ln [N(J)/g(J)]$ {\it vs.\/} $E_J$, excluding $J=4$.}
\end{deluxetable}

\subsection{Unresolved Saturated Substructures in Component
3}\label{unres_sat}

For the right-hand peaks (Component 3) in $J$ = 0, 1 and 2, the weakest
transitions show more H$_2$ than indicated in Figs.~\ref{j0fig} to
\ref{j2fig}, which are based on generally much stronger transitions. 
This behavior reveals the presence of very narrow substructures in
Component 3 that are saturated and not resolved by the instrument. 
Jenkins \markcite{3184} (1996) has shown how one may take any pair of
lines (of different strength) that show a discrepancy in their values of
$N_a(v)$, as evaluated from Eqs.~\ref{tau_a} and \ref{N_a}, and evaluate
a correction to $\tau_a(v)$ of the weaker line that compensates for the
under-representation of the smoothed real optical depths $\tau(v)$.  In
effect, this correction is a method of extrapolating the two distorted
$N_a(v)$'s to a profile that one would expect to see if the line's
transition strength was so low that the unresolved structures had their
maximum (unsmoothed) $\tau (v)\ll 1$.

Unfortunately, we found that for each of the three lowest $J$ levels,
different pairs of lines yielded inconsistent results.  In each case, an
application of the analysis of the first and second weakest lines gave
column densities considerably larger than the same procedure applied to
the second and third weakest lines.  We list below a number of
conjectures about the possible cause(s) for this effect:
\begin{enumerate}
\item The functional forms of the distributions of subcomponent
amplitudes and velocity widths are so bizarre, and other conditions are
exactly right, that the assumptions behind the workings of the
correction procedure are not valid.  As outlined by Jenkins
\markcite{1355,3184} (1986, 1996), these distributions would need to be
very badly behaved.
\item We have underestimated the magnitudes of the errors in the
determinations of scattered light in the spectrum, which then reflect on
the true levels of the zero-intensity baselines and, consequently, the
values of $\tau_a(v)$ near maximum absorption.
\item The transition $f$-values that we have adopted are wrong.  The
sense of the error would be such that the weakest lines are actually
somewhat stronger than assumed, relative to the $f$-values of the next
two stronger lines.  Another alternative is that the second and third
strongest lines are much closer together in their $f$-values than those
that were adopted.
\end{enumerate}

While we can not rigorously rule out possibilities (1) and (2) above, we
feel that they are unlikely to apply.  Regarding possibility (3), the
$f$-values are the product of theoretical calculations, and to our
knowledge only some of the stronger transitions have been verified
experimentally \markcite{3186} (Liu et al. 1995).  It is interesting to
see if there is any observational evidence outside of the results
reported here that might back up the notion that alternative (3) is the
correct explanation.

We are aware of two potentially useful examples where the weakest
members of the Lyman series have been seen in the spectra of
astronomical sources.  One is in a survey of many stars by Spitzer,
Cochran \& Hirshfeld \markcite{1015} (1974),\footnote{There are many
papers that report observations of H$_2$ made by the {\it Copernicus}
satellite.  Oddly enough, the paper by Spitzer  et al. \markcite{1015}
(1974) is the only one that includes measurements of the weakest lines.}
and another is an array of H$_2$ absorption features identified by
Levshakov \& Varshalovich \markcite{3318} (1985) and Foltz, et al.
\markcite{1078} (1988) at $z$ = 2.811 in the spectrum of the quasar
PKS~0528$-$250.  The quasar absorption lines have subsequently been
observed at much higher resolution by Songaila \& Cowie \markcite{287}
(1995) using the Keck Telescope.

In the survey of Spitzer  et al. \markcite{1015} (1974), the only target
that showed lines from $J$=0 that were not on or very close to the flat
portion of the curve of growth (or had an uncertain measurement of the
Lyman 0$-$0\,R(0) line) was 30 CMa.  The 10m\AA\ equivalent width
measured for this line is above a downward extrapolation of the the
trend from the stronger lines.  If the line's value of $\log f\lambda$
were raised by 0.28 in relation with the others, the measured line
strength would fall on their adopted curve of growth.  Unfortunately, we
can not apply the same test for the Lyman 0$-$0\,P(1) or
0$-$0\,R(2)lines, the two weakest lines that we could use here for the
next higher $J$ levels, because these lines were not observed by Spitzer
et al.
 
The H$_2$ lines that appear in the spectrum PKS~0528$-$250 are created
by a heavy-element gas system that is moving at only 2000 to 3000
km~s$^{-1}$ with respect to the quasar (and hence one that is not very
far away from the quasar).  The overall widths of the H$_2$ lines of
about 20~km~s$^{-1}$ were resolved in the R = 36,000 spectrum of
Songaila \& Cowie \markcite{287} (1995), but the shallow Lyman 0$-$0,
1$-$0 and 2$-$0\,R(0) features showed a strengthening that was far less
than the changes in their relative $f$-values.  Songaila \& Cowie
interpreted this behavior as the result of saturation in the lines if
they consisted of a clump of 5 unresolved, very narrow features, each
with $b$ = 1.5 km~s$^{-1}$, distributed over the observed velocity
extent of the absorption.  One might question how plausible it is to
find gas clouds with such a small velocity dispersion that could cover a
significant fraction the large physical dimension of the
continuum-emitting region of the quasar.  As an alternative, we might
accept the notion that the lines do not contain unresolved saturated
components, but instead, that the real change in the $f$-values is less
than assumed.

Finally, we turn to our own observations.  In our recording of the Lyman
0$-$0\,R(0) line in our spectrum of $\zeta$~Ori, the amplitude of the
$\tau_a(v)$ profile of Component~1 (about $4\sigma$ above the noise), in
relation to that of Component~3, is not much different than what may be
seen in the next stronger line, 1$-$0\,R(0).  If significant distortion
caused by unresolved, saturated substructures were occurring for
Component 3 in the latter, the size difference for the two components
would be diminished, contrary to what we see in the data.  If one were
to say that the difference in $\log (f\lambda)$ for the two lines were
smaller by 0.4, we would obtain $N_a(v)$'s that were consistent with
each other.

We regard the evidence cited above as suggestive, but certainly not
conclusive, evidence that our problems with the disparity of answers for
$N_a(v)$ might be caused by incorrect relative $f$-values.  Even if this
conjecture is correct, we still do not know whether the stronger or
weaker $f$-values need to be revised.  In view these uncertainties, we
chose to derive $N(v)$ for Component~3 by two different methods,
Method~A and Method~B, outlined in the following two subsections. Total
column densities $\int N(v)\,dv$ derived each way are listed in
Table~\ref{comp_summary}. 

\subsubsection{Method A}\label{method_A}

Method~A invokes the working assumption that the adopted $f$-value for
the weakest line is about right, and that there is a problem with the
somewhat stronger lines.  If this is correct, then our only recourse is
to derive $N_a(v)$ from this one line through the use of
Eqs.~\ref{tau_a} and \ref{N_a} and assume that the correction for
unresolved saturated substructures is small.  For $J=0$, 1 and 2, we
used the Lyman 0$-$0\,R(0), 0$-$0\,P(1) and 0$-$0\,R(2) lines,
respectively.  (The weakest line for $J=2$, 0$-$0\,P(2) could not be
used; see note $b$ of Table~\ref{j2table}.)

\subsubsection{Method B}\label{method_B}

Here we assume that the published $f$-value for the weakest line is too
small, but that the values for the next two stronger lines are correct. 
We then derive corrections for $\tau_a(v)$ for the weaker line using the
method of Jenkins \markcite{3184} (1996).  While the errors in this
extrapolation method can be large, especially after one considers the
effects of the systematic deviations discussed earlier [items (2) and
(3) covered in \S\ref{line_meas}], under the present circumstances they
are probably not much worse than the arbitrariness in the choice of
whether Method~B is any better than Method~A or some other way to derive
$N_a(v)$.  Lyman band line pairs used for this method were 1$-$0\,R(0)
and 2$-$0\,R(0) for $J=0$, 0$-$0\,R(1) and 1$-$0\,P(1) for $J=1$, and
1$-$0\,P(2) and 1$-$0\,R(2) for $J=2$.

\subsection{Profile Changes with $J$ for
Component~1}\label{prof_changes}

\begin{figure}
\plotone{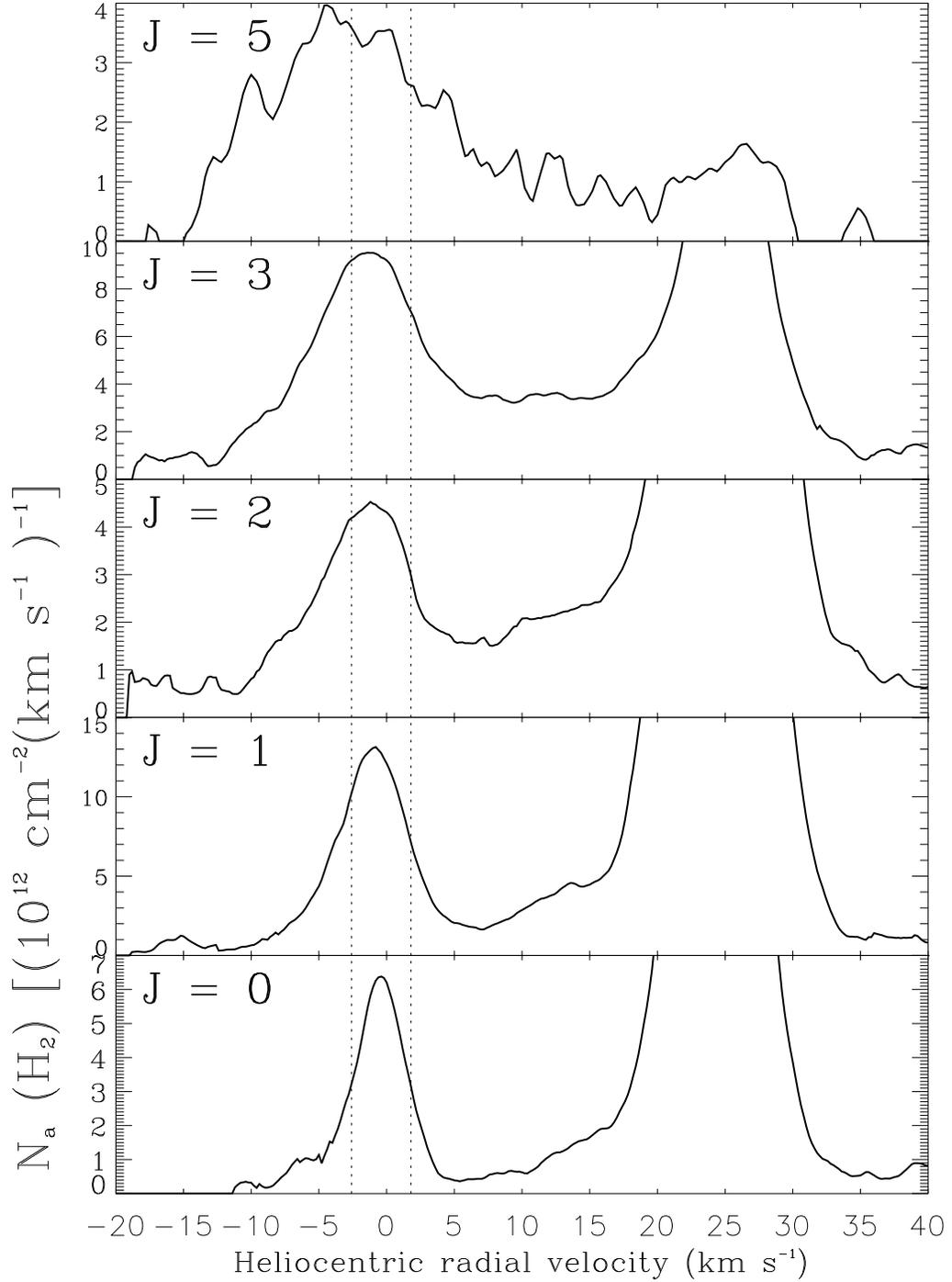}
\caption{Plots of $\log N_a(v)$ versus $v$ scaled such that
the heights of the peaks for Component~1 are nearly identical.  To
facilitate comparisons of widths and velocity centers across different
$J$ levels, the two vertical, dotted lines mark the half amplitude
points of the $J=0$ profile.\label{j0-j5fig}}
\end{figure}

Figures \ref{j0fig} to \ref{j5fig} show very clearly that the profiles
for Component~1 have widths that progressively increase as the
rotational quantum numbers go from $J=0$ to $J=5$. 
Figure~\ref{j0-j5fig} shows a consolidation of the results from
Figs.~\ref{j0fig} to \ref{j5fig}: the valleys of $\chi^2-\chi^2_{\rm
min}$ are depicted as lines [now in a linear representation for
$N_a({\rm H}_2)$], and the profiles are stacked vertically to make
comparisons for different $J$ in Component~1 more clear.  In addition to
showing the changes in profile widths, this figure also shows that there
is a small ($\sim 1~{\rm km~s}^{-1}$) shift toward negative velocities
with increasing $J$ up to $J=3$, followed by a more substantial shift
for $J=5$.

\placefigure{j0-j5fig}

A simple, approximate way to express numerically the information shown
in Fig.~\ref{j0-j5fig} is to assume that most of the H$_2$ at each $J$
level has a one-dimensional distribution of velocity that is a Gaussian
function characterized by a peak value for $N(v)$, $N_{\rm max}$, a
central velocity, $v_0$, and a dispersion parameter, $b$.  We can then
ascertain what combinations of these 3 parameters give an acceptable fit
to the data as defined, for example, by values $\chi^2-\chi^2_{\rm min}
< 7.8$ that lead to a 95\% confidence limit.  We carried out this study
with $\chi^2$'s, of the type displayed in Figs.~\ref{j0fig} to
\ref{j5fig}, summed over velocity points spaced 1.6~km~s$^{-1}$ apart to
assure statistical independence.  Table~\ref{chi2} summarizes the
results of that investigation.  The quantities $v_{\rm min}$ and $v_{\rm
max}$ are the velocity limits over which the fits were evaluated.  The
error bounds are defined only by the $\chi^2$ limits and do not include
systematic errors, such as those that arise from errors in $f$-values or
our overall adopted zero-point reference for radial velocities.  For
given $J$ levels, there are small differences between the preferred
$\log (N_{\rm max}\sqrt{\pi}b)$ and the log column densities given in
Table~\ref{comp_summary} caused by real departures from the Gaussian
approximations ($J=5$ shows the largest deviation, 0.08 dex, as one
would expect from the asymmetrical appearance shown in
Fig.~\ref{j0-j5fig}).

\placetable{chi2}
\begin{deluxetable}{
c    % descr.
c    % J=0
c    % J=1
c    % J=2
c    % J=3
c    % J=5
}

\tablewidth{0pt}
\tablecaption{Gaussian Fits to Component~1\label{chi2}}
\tablehead{
\colhead{} &
\colhead{$J=0$} &
\colhead{$J=1$} &
\colhead{$J=2$} &
\colhead{$J=3$} &
\colhead{$J=5$}
}
\startdata
$v_{\rm max}$ (km~s$^{-1}$)&+4&+4&+5&+5&+7\nl
$v_{\rm min}$&$-$5&$-$6&$-$8&$-$8&$-$14\nl
\tablevspace{15pt}
Largest $\log N_{\rm max}~[{\rm cm}^{-2}({\rm
km~s}^{-1})^{-1}]$&12.82&13.12&12.68&13.00&12.66\nl
Most probable $\log N_{\rm max}$&12.77&13.09&12.62&12.96&12.56\nl
Smallest $\log N_{\rm max}$&12.72&13.06&12.58&12.94&12.46\nl
\tablevspace{15pt}
Largest $v_0$
(km~s$^{-1}$)\tablenotemark{a}&$-$0.3&$-$0.9&$-$1.0&$-$1.0&$-$1.0\nl
Most probable $v_0$&$-$0.5&$-$1.0&$-$1.5&$-$1.3&$-$2.9\nl
Smallest $v_0$&$-$0.7&$-$1.2&$-$2.0&$-$1.6&$-$4.4\nl
\tablevspace{15pt}
Largest $b$ (km~s$^{-1}$)\tablenotemark{b}&3.2&4.2&7.0&6.8&14\nl
Most probable $b$&2.9&3.9&6.0&6.5&9.4\nl
Smallest $b$&2.6&3.8&5.2&6.0&7.2\nl
\enddata
\tablenotetext{a}{Heliocentric radal velocity of the profile's center.}
\tablenotetext{b}{Includes instrumental broadening and registration
errors (see \S\protect\ref{prof_changes}).  Hence, the real $b$ should
equal about $\sqrt{b_{\rm obs}^2-(2.8~{\rm km~s}^{-1})^2}$.}
\end{deluxetable}

To determine the real widths of the profiles, one must subtract in
quadrature two sources of broadening in the observations.  First, there
is the instrumental broadening of each line in the spectrum that we
recorded, as discussed in \S\ref{wl_scale}.  Adding to this effect are
the small errors in registration of the lines, as they are combined to
create the $\chi^2-\chi^2_{\rm min}$ plots (Figs.~\ref{j0fig} to
\ref{j5fig}).  From the apparent dispersion of line centers at a given
$J$, we estimate the rms registration error to be 0.5~km~s$^{-1}$.  We
estimate that the effective $b$ parameter for these two effects combined
should be about 2.8~km~s$^{-1}$, and thus the formula given in note $a$
of Table~\ref{chi2} should be applied to obtain a best estimate for the
true $b$ of each H$_2$ profile (the results for the lowest $J$ levels
will not be very accurate, since $b_{\rm obs}$ is only slightly greater
than 2.8~km~s$^{-1}$).

The results shown in Fig.~\ref{j0-j5fig} and Table~\ref{chi2} show two
distinct trends of the profiles with increasing $J$.  First, the most
probable values for the widths $b$ increase in a steady progression from
$J=0$ to $J=5$.  Second, the most probable central velocities $v_0$
become steadily more negative with increasing $J$, except for an
apparent reversal between $J=2$ and $J=3$ that is much smaller than our
errors.  It is hard to imagine that systematic errors in the
observations could result in these trends.  The absorption lines for
different $J$ levels appear in random locations in the spectral image
formats, so any changes in the spectral resolution or distortions in our
wavelength scale should affect all $J$ levels almost equally.  

\section{Discussion}\label{discussion}
\subsection{Preliminary Remarks}\label{prelim}

The information given in Table~\ref{comp_summary} shows that the 3
molecular hydrogen velocity components toward $\zeta$~Ori~A have
populations in different $J$ levels that, to a reasonable approximation,
conform to a single rotational excitation temperature in each case. 
This behavior seems to reflect what has been observed elsewhere in the
diffuse interstellar medium.  For instance, in their survey of 28 lines
of sight, Spitzer,  et al. \markcite{1015} (1974) found that for
components that had $N(J=0)\lesssim 10^{15}{\rm cm}^{-2}$, a single
excitation temperature gave a satisfactory fit to all of the observable
$J$ levels.  By contrast, one generally finds for much higher column
densities that there is bifurcation to two temperatures, depending on
the $J$ levels [see, e.g., Fig.~2 of Spitzer \& Cochran \markcite{1212}
(1973)].  This is a consequence of the local density being high enough
to insure that collisions dominate over radiative processes at low to
intermediate $J$ and thus couple the level populations to the local
kinetic temperature, whereas for higher $J$ the optical pumping can take
over and yield a somewhat higher temperature.  For cases where the total
column densities are exceptionally low [$N({\rm H}_2)\approx 10^{13}{\rm
cm}^{-2}$ for such stars as $\zeta$~Pup, $\gamma^2$~Vel and $\tau$~Sco],
the rotation temperatures can be as high as about 1000K.  This behavior
is consistent with what we found for our Components~1 and 2.  Our
Component~3 has a somewhat lower excitation temperature, but one that is
in accord with other lines of sight that have $N({\rm H}_2)\approx
10^{15}{\rm cm}^{-2}$ in the sample of Spitzer,  et al. \markcite{1015}
(1974).

It is when we go beyond the information conveyed by just the column
densities and study changes in the profiles for different $J$ that we
uncover some unusual behavior.  Here, we focus on Component~1, where the
widths and velocity centroids show clear, progressive changes with
rotational excitation.  While Component~3 also shows some broadening
with increasing $J$, the magnitude of the effect is less, and it is
harder to quantify because there are probably unresolved, saturated
structures that distort the $N_a(v)$ profiles.  The changes in
broadening with $J$ are inconsistent with a simple picture that, for the
most diffuse clouds, the excitation of molecular hydrogen is caused by
optical pumping out of primarily the $J=0$ and 1 levels by uv starlight
photons in an optically thin medium.

We might momentarily consider an explanation where the strength of the
optical pumping could change with velocity, by virtue of some shielding
in the cores of some of the strongest pumping lines.  However, in the
simplest case we can envision, one where the light from $\zeta$~Ori
dominates in the pumping, the shielding is not strong enough to make
this effect work.  For example, in Component~1 we found $\log N({\rm
H}_2)=13.53$ (Table~\ref{comp_summary}) and a largest possible {\it real
value\/}\footnote{See note $b$ of Table~\protect\ref{chi2}} of
$b=1.55~{\rm km~s}^{-1}$ for molecules in the $J=0$ level.  We would
need to have a pumping line from $J=0$ with a characteristic strength
$\log f\lambda=3.0$ to create an absorption profile $1-I(v)/I_0$ that is
saturated enough to have it appear, after a convolution with our
instrumental profile, as broad as the observed $N_a(v)$ for molecules in
the $J=2$ state.\footnote{This simple proof is a conservative one, since
it neglects other processes that tend to make the $J=2$ profile as
narrow as that for $J=0$, such as pumping from many other, much weaker
lines or the coupling of molecules in the $J=2$ state with the kinetic
motions of the gas through elastic collisions.}  In reality, the
strongest lines out of $J=0$ have $\log f\lambda$ only slightly greater
than 1.8 (see Table~\ref{j0table}).  Likewise, the width of the $N_a(v)$
profile for $J=3$ can only be matched with a pumping line out of $J=1$
with $\log f\lambda=2.0$, again a value that is much higher than any of
the actual lines out of this level (see Table~\ref{j1table}). Thus, if
we are to hold on to the notion that line shielding could be an
important mechanism, we must abandon the idea that $\zeta$~Ori is the
source of pumping photons.

We could, of course, adopt a more imaginative approach and propose that
light from another star is responsible for the pumping.  Then, we could
envision that a significant concentration of H$_2$ just off our line of
sight could be shielding (at selective velocities) the radiation for the
molecules that we can observe.  While this could conceivably explain why
the profiles for $J>1$ look different from those of $J=0$ or 1, it does
not address the problem that the profile for $J=1$ disagrees with that
of $J=0$. (The coupling of these two levels by optical pumping is very
weak.)  As indicated by the numbers in Table~\ref{chi2}, both the
velocity widths and their centroids for these lowest two levels differ
by more than the measurement errors.

Another means for achieving a significant amount of rotational
excitation is heating due to the passage of a shock --- one that is slow
enough not to destroy the H$_2$ \markcite{2812} (Aannestad \& Field
1973). Superficially, we might have imagined that Component~1 is a
shocked portion of the gas that was originally in Component~3, but that
is now moving more toward us, relatively speaking.  However this picture
is in conflict with the change in velocity centroids with $J$, for the
gas would be expected to speed up as it cools in the postshock zone
where radiative cooling occurs.  Our observations indicate that the
cooler (rear) part of this zone that should emphasize the lower $J$
levels is actually traveling more slowly.

From the above argument on the velocity shift, it is clear that if we
are to invoke a shock as the explanation for the profile changes, we
must consider one that is headed in a direction away from us.  If this
is so, we run into the problem that we are unable to see any H$_2$ ahead
of this shock, i.e, at velocities more negative than Component~1.  Thus,
instead of creating a picture where existing molecules are accelerated
and heated by a shock, we must turn to the idea that perhaps the
molecules are formed for the first time in the dense, compressed
postshock zone, out of what was originally atomic gas undergoing cooling
and recombination.  In this case, one would look for a shock velocity
that is relatively large, so that the compression is sufficient to raise
the density to a level where molecules can be formed at a fast rate.

\subsection{Evidence of a Shock that could be Forming
H$_2$}\label{shock}

\subsubsection{Preshock Gas}\label{preshock_gas}

There is some independent evidence from atomic absorption lines that we
could be viewing a bow shock created by the obstruction of a flow of
high velocity gas coming toward us, perhaps a stellar wind or a
wind-driven shell \markcite{1696} (Weaver et al. 1977).  A reasonable
candidate for this obstruction is a cloud that is responsible for the
low-ionization atomic features that can be seen near $v=0~{\rm
km~s}^{-1}$.

In the IMAPS spectrum of $\zeta$~Ori~A, there are some strong
transitions of C~II (1036.337\AA) and N~II (1083.990\AA) that show
absorption peaks at $-$94~km~s$^{-1}$, plus a smaller amount of material
at slightly lower velocities \markcite{2956} (Jenkins 1995).  Features
from doubly ionized species are also present at about the same velocity,
i.e., C~III, N~III, Si~III, S~III \markcite{1151} (Cowie, Songaila, \&
York 1979) and Al~III (medium resolution GHRS spectrum in the HST
archive\footnote{Exposure identification: Z165040DM.}).  Absorption by
strong transitions of O~I and N~I are not seen at $-$94~km~s$^{-1}$
however.  The moderately high state of ionization of this rapidly moving
gas, a condition similar to that found for high velocity gas in front of
23~Ori by Trapero  et al. \markcite{356} (1996), may result from either
photoionization by uv radiation from the Orion stars or collisional
ionization at a temperature somewhat greater than $10^4$K.  

\begin{figure}
\plotone{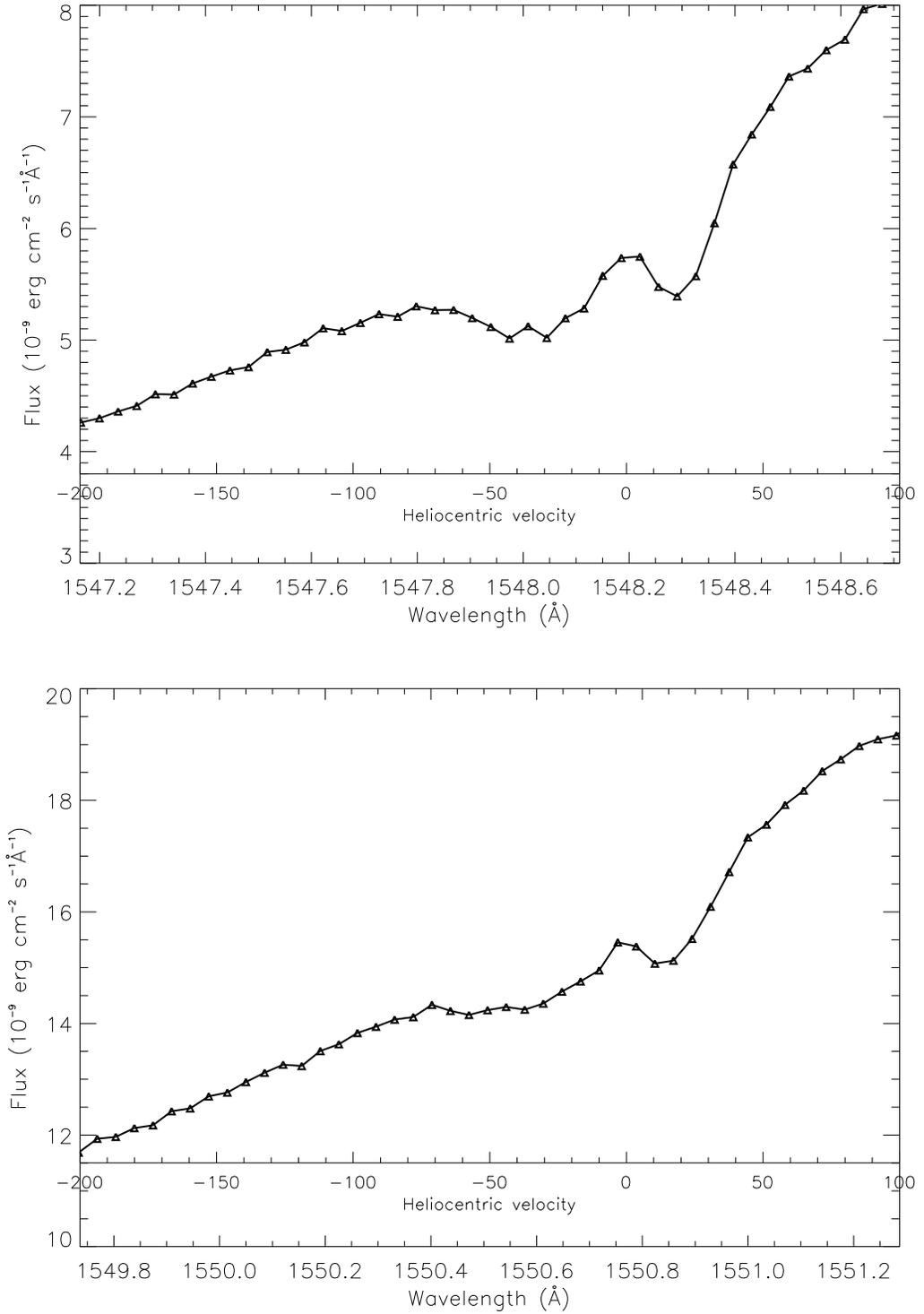}
\caption{A medium resolution (R = 20,000) recording of the
C~IV doublet in a spectrum of $\zeta$~Ori~A taken with the G160M grating
of GHRS on the Hubble Space Telescope.  A correction of +0.06\AA\ has
been added to the wavelength scale of calibrated file for exposure
Z1650307T to reflect the offset in a 1554.9285\AA\ calibration line that
may be seen in exposure Z1650306T.\label{CIV}}
\end{figure}

Figure~\ref{CIV} shows spectra that we recovered from the HST
archive\footnote{Again, a medium resolution GHRS spectrum: Exposure
identification: Z1650307T.} in the vicinity of the C~IV doublet (1548.2,
1550.8\AA).  We determined an upper limit $\log N({\rm C~IV})<12.2$ at
$v\approx -90~{\rm km~s}^{-1}$.  When this result is compared with the
determination $\log N({\rm C~II})=13.84$ \markcite{1151} (Cowie,
Songaila, \& York 1979) or 13.82 (IMAPS spectrum), we find that
$T<20,000$K if we use the collisional ionization curves of Benjamin \&
Shapiro \markcite{327} (1996) for a gas that is cooling isobarically. 
(A similar argument arises from an upper limit for N~V/N~II, but the
resulting constraint on the temperature is weaker.)  There is
considerably more Si~III than Si~II in the high velocity gas
\markcite{1151} (Cowie, Songaila, \& York 1979), but this is may be due
to photoionization.  Thus, to derive a lower limit for the temperature
of the gas, we must use a typical equilibrium temperature for an H~II
region, somewhere in the range $8,000 < T < 12,000$K \markcite{1855}
(Osterbrock 1989).

\placefigure{CIV}

\subsubsection{Immediate Postshock Gas}\label{immediate_PS}

Figure~\ref{CIV} shows that there is a broad absorption from C~IV
centered at a velocity of about $-36~{\rm km~s}^{-1}$, in addition to a
narrower peak at about +20~km~s$^{-1}$.  The equivalent widths of 50 and
22~m\AA\ for the broad, negative velocity components for the transitions
at 1548.2 and 1550.8\AA, respectively, indicate that $\log N({\rm
C~IV})=13.1$, a value that is in conflict with the upper limit $\log
N({\rm C~IV})=11.9$ obtained by Cowie,  et al.  \markcite{1151} (1979). 
Absorptions by Si~III (1206.5\AA) and Al~III (1854.7\AA) are also
evident at $-20$ and $-15~{\rm km~s}^{-1}$,
respectively.\footnote{Archive exposure identifications: Z165040CM and
Z165040DM.}  We propose that these high ionization components arise from
collisionally ionized gas behind the shock front.  (Ultraviolet
radiation from the shock front also helps to increase the ionization of
the downstream gas.)  The width of the C~IV feature shown in
Fig.~\ref{CIV} reflects the effects of thermal doppler broadening,
instrumental smearing, and the change in velocity as the gas cools to
the lowest temperature that holds any appreciable C~IV.

\subsubsection{Properties of the Shock}\label{shock_properties}

We return to our conjecture that the preshock gas flow is being
intercepted by an obstacle at $v\approx 0~{\rm km~s}^{-1}$, and thus the
front itself is at this velocity.  (While this assumption is not backed
up by independent evidence, it is nevertheless a basic premise behind
our relating the atomic absorption line data to our interpretation in
\S\ref{warm_formation} and \S\ref{cool_formation} of how the H$_2$ in
Component~1 is formed in a region where there is a large compression and
a temperature that is considerably lower than that of the immediate
postshock gas.)  The fact that the C~IV feature does not appear at
\onequarter\ times that of the high velocity (preshock) C~II and N~II
features indicates that the compression ratio is less than the value 4.0
for strong shocks with an adiabatic index $\gamma=5/3$.  This is
probably a consequence of either the ordinary or Alfv\'en Mach numbers
(or both) not being very high.  For example, if the preshock magnetic
field and density were 5$\mu$G and $n_0=0.1~{\rm cm}^{-3}$, the Alfv\'en
speed would be 29~km~s$^{-1}$.  For $T=20,000$K, the ordinary sound
speed would be 21~km~s$^{-1}$, and under these conditions the
compression ratio would be only 2.67 [cf. Eq.~2.19 of Draine \& McKee
\markcite{187} (1993)] if the magnetic field lines are perpendicular to
the shock normal.  This value is close to the ratio of velocities of the
preshock and postshock components, $(-94~{\rm km~s}^{-1})/(-36~{\rm
km~s}^{-1})=2.6$.  The immediate postshock temperature would be about
$2.3\times 10^5$K.

Our simple picture of a shock that is moderated by a transverse magnetic
field adequately explains the velocity difference between the two atomic
components, but it fails when we try to fit the kinematics of the much
cooler gas where we find H$_2$.  If we follow the material in the
postshock flow to the point that radiative cooling has lowered the
temperature to that of the preshock gas or below, we expect to have a
final compression ratio equal to 3.7, i.e., the number that we would
expect for an ``isothermal shock'' [cf. Eq.~2.27 of Draine \& McKee
\markcite{187} (1993)].  This limited amount of compression would mean
that the cool, H$_2$-bearing gas would appear at a velocity of
$(-94~{\rm km~s}^{-1})/3.7$ = $-25~{\rm km~s}^{-1}$, a value that is
clearly inconsistent with what we observe.  

A resolution of the inconsistency between the kinematics noted above and
the theoretical picture of a shock dominated by magnetic pressure could
be obtained if, instead of having the initial magnetic field lines
perpendicular to the shock normal, the field orientation is nearly
parallel to the direction of the flow.  (Intuitively, this arrangement
seems more plausible, since the field lines are likely to be dragged
along by the gas.)  The picture than can then evolve to the more complex
situation where there is a ``switch-on'' shock, giving an initial
moderate compression and a sudden deflection of the velocity flow and
direction of the field lines.  As described by Spitzer \markcite{2832,
33} (1990a, b), this phase may then be followed by a downstream
``switch-off'' shock that redirects the flow and field lines to be
perpendicular to the front and allows further compression of the gas up
to values equal to the square of the shock's ordinary Mach number, i.e.,
the compression produced by a strong shock without a magnetic field.

In order to obtain a solution for a switch-on shock, one must satisfy
the constraint that the Alfv\'en speed must be greater than slightly
more than half of the shock speed [cf. Eq. 2.21 of Draine \& McKee
\markcite{187} (1993)].  Thus, we must at least double the Alfv\'en
speed of the previous example by either raising the preshock magnetic
field, lowering the density, or both.  If this speed equalled $58~{\rm
km~s}^{-1}$, the compression ratio in the switch-on region should be
$(94/58)^2=2.6$, i.e., the square of the Alfv\'en Mach number, a value
that is again very close to our observed ratio of gas velocities on
either side of the front. [There is a complication in deriving a
compression ratio from an observation taken at some arbitrary viewing
direction through a switch-on shock.  Behind the front, the gas acquires
a velocity vector component that is parallel to the front.  For an
inclined line of sight, this component can either add to or subtract
from the projection of the component perpendicular to the front, which
is the quantity that must be compared to the (again projected) preshock
velocity vector when one wants to obtain a compression ratio.  However
in our situation it seems reasonable to suppose that a wind from
$\zeta$~Ori is the ultimate source of high velocity gas, and this in
turn implies that the shock front is likely to be nearly perpendicular
to the line of sight.]  While this picture is still rather speculative,
we will adopt the view that, through the mechanism of the switch-off
shock, the magnetic fields do not play a significant role in limiting
the amount of compression at the low temperatures where H$_2$ could
form.

One additional piece of information is a limit on the preshock density
$n_0$. Cowie,  et al. \markcite{1151} (1979) obtained an upper limit for
the electron density $n_{\rm e}<0.3~{\rm cm}^{-3}$ from the lack of a
detectable absorption feature from C~II in an excited fine-structure
level (assuming $T=10^4$K).  Since there is virtually no absorption seen
for lines of N~I or O~I at the high velocities in front of the shock, we
can be confident that the hydrogen is almost fully ionized and thus the
limit for $n_{\rm e}$ applies to the total density.  For the purposes of
argument in the discussions that follow, we shall adopt a value
$n_0=0.1~{\rm cm}^{-3}$, as we have done earlier.

\subsection{Formation of H$_2$ in a Warm Zone}\label{warm_formation}
\subsubsection{Reactions and their Rate Constants}\label{reactions}

In the light of evidence from the atomic lines that a standing shock may
be present, we move on to explore in a semiquantitative way the
prospects that H$_2$ forming behind this front could explain our
observations.  For several reasons, we expect that an initial zone where
$T \gtrsim 10^4$K will produce no appreciable H$_2$.  At these
temperatures the gas is mostly ionized, and for $T > 18,000$K collisions
with electrons will dissociate H$_2$ very rapidly (Draine \& Bertoldi,
in preparation).  Furthermore, the column density of material at
$T>6500$K is not large because the cooling rate is high.  As soon as the
gas has reached 6500K, there is a significant, abrupt reduction in the
cooling rate while there is still some heating of the gas by ionizing
radiation produced by the much hotter, upstream material.  These effects
create a plateau in the general decrease of temperature with postshock
distance [see Fig.~3 of Shull \& McKee \markcite{1800} (1979)].

The 6500K plateau, extending over a length of approximately $2\times
10^{16}n_0^{-1}{\rm cm}$, seems to be a favorable location for
synthesizing the initial contribution of H$_2$ that we could be viewing
in the upper $J$ levels.  Its velocity with respect to much cooler gas
should be about $(-36~{\rm km~s}^{-1})\times (6500{\rm K})/(2.3\times
10^5{\rm K})=-1.0~{\rm km~s}^{-1}$ if the conditions are approximately
isobaric. This velocity difference is consistent, to within the
observational errors, with the shift between the peaks at $J=5$ and
$J=0$, with the latter emphasizing molecules in the material that has
cooled much further and come nearly to a halt.  Considering that the
fractional ionization over the temperature range $6500 > T > 2000$K is
$0.5\gtrsim n_{\rm e}/n_{\rm H}\gtrsim 0.03$ \markcite{1800} (Shull \&
McKee 1979), we anticipate that potentially important sources of H$_2$
arise from either the formation of a negative hydrogen ion,
\begin{eqnarray}\label{c8}
{\rm H} + e&\rightarrow &{\rm H}^- + h\nu\nonumber\\
C_{\ref{c8}}&=&1.0\times 10^{-15}T_3\exp(-T_3/7)~{\rm cm}^3{\rm s}^{-1}
\end{eqnarray}
($T_3$ is the gas's temperature in units of $10^3$K) followed by the
associative detachment,
\begin{eqnarray}\label{c9}
{\rm H}^- + {\rm H}&\rightarrow &{\rm H}_2 + e\nonumber\\
C_{\ref{c9}}&=&1.3\times 10^{-9}{\rm cm}^3{\rm s}^{-1}
\end{eqnarray}
or the production of H$_2^+$ by radiative association,
\begin{eqnarray}\label{c2}
{\rm H} + {\rm H}^+&\rightarrow &{\rm H}_2^+ + h\nu\nonumber\\
C_{\ref{c2}}&=&4.1\times 10^{-17}{\rm cm}^3{\rm s}^{-1}
\end{eqnarray}
followed by its reaction with neutral atoms,
\begin{eqnarray}\label{c3}
{\rm H}_2^+ + {\rm H}&\rightarrow &{\rm H}_2 + {\rm H}^+\nonumber\\
C_{\ref{c3}}&=&1.0\times 10^{-10}{\rm cm}^3{\rm s}^{-1}
\end{eqnarray}
\markcite{2611, 3272} (Black 1978; Black, Porter, \& Dalgarno 1981). 
The rate constants for the above reactions (plus the destruction
reactions \ref{c10} and \ref{c5} below) are the same as those adopted by
Culhane \& McCray \markcite{3157} (1995) in their study of H$_2$
production in a supernova envelope.  Later, as the gas becomes cooler,
denser and mostly neutral, we expect that the formation of H$_2$ on the
surfaces of dust grains,
\begin{eqnarray}\label{cg}
2{\rm H} + {\rm grain}&\rightarrow & {\rm H}_2 + {\rm grain}\nonumber\\
\lbrack {\rm applicable~to}~n({\rm H})^2\rbrack
~~C_{\ref{cg}}&=&{10^{-16}T_3^{0.5}\over
1+1.3T_3^{0.5}+2T_3+8T_3^2}~{\rm cm}^3{\rm s}^{-1}
\end{eqnarray}
should start to become more important \markcite{3190} (Hollenbach \&
McKee 1979).  We will address this possibility in
\S\ref{cool_formation}.

In order to evaluate the effectiveness of reactions \ref{c9} and
\ref{c3} in producing H$_2$ in the warm gas, we must consider the most
important destruction processes that counteract the production of the
feedstocks H$^-$ (reaction~\ref{c8}) and H$_2^+$ (reaction~\ref{c2}). 
Radiative dissociation of H$^-$ by uv starlight photons (i.e., the
reverse of reaction~\ref{c8}),
\begin{eqnarray}\label{c8-1}
{\rm H}^- + h\nu&\rightarrow & {\rm H} + e\nonumber\\
\beta_{\ref{c8-1}}&=& 1.9\times 10^{-7}{\rm s}^{-1}
\end{eqnarray}
is generally the most important mechanism for limiting the eventual
production of H$_2$ in partially ionized regions of the interstellar
medium.  The value for $\beta_{\ref{c8-1}}$ is adopted from an estimate
for this rate of destruction in our part of the Galaxy by Fitzpatrick \&
Spitzer \markcite{2701} (1994).  Less important ways of destroying H$^-$
include recombination with protons,
\begin{eqnarray}\label{c10}
{\rm H}^- + {\rm H}^+&\rightarrow &2{\rm H}\nonumber\\
C_{\ref{c10}}&=&7\times 10^{-8}T_3^{-0.4}{\rm cm}^3{\rm s}^{-1}
\end{eqnarray}
and, of course, the production of H$_2$ (reaction~\ref{c9}).  H$_2^+$ is
destroyed by the reaction with electrons,
\begin{eqnarray}\label{c5}
{\rm H}_2^+ + e&\rightarrow &2{\rm H}\nonumber\\
C_{\ref{c5}}&=& 1.4\times 10^{-7}T_3^{-0.4}{\rm cm}^3{\rm s}^{-1}
\end{eqnarray}
and the creation of H$_2$ in reaction~\ref{c3}.  We can safely disregard
the interaction of H$_2^+$ with H$_2$,
\begin{eqnarray}\label{c4}
{\rm H}_2^+ + {\rm H}_2&\rightarrow &{\rm H}_3^+ + {\rm H}\nonumber\\
C_{\ref{c4}}&=& 2.1\times 10^{-9}{\rm cm}^3{\rm s}^{-1}
\end{eqnarray}
because $C_{\ref{c4}}n({\rm H}_2)\ll C_{\ref{c5}}n(e)+C_{\ref{c3}}n(H)$. 
Finally, our end product H$_2$ is destroyed by photodissociation,
\begin{eqnarray}\label{c17}
{\rm H}_2 + h\nu&\rightarrow & 2{\rm H}\nonumber\\
({\rm optically~thin})~~\beta_{\ref{c17}}&=&3.4\times 10^{-11}{\rm
s}^{-1}~.
\end{eqnarray}
Our adopted general value for $\beta_{\ref{c17}}$ makes use of Jura's
\markcite{1276} (1974) calculation of $\beta_{\ref{c17}}=5.4\times
10^{-11}{\rm s}^{-1}$ for a flux of $4\pi J_\lambda=2.4\times
10^{-6}{\rm erg~cm}^{-2}{\rm s}^{-1}{\rm \AA}^{-1}$ at 1000\AA, but
rescaled to a local flux of $1.5\times 10^{-6}{\rm erg~cm}^{-2}{\rm
s}^{-1}{\rm \AA}^{-1}$ calculated by Mezger, et al. \markcite{3273}
(1982).  A large fraction of this background may come from sources that
are behind or within cloud complexes containing H$_2$.  If this is true,
the stellar radiation in the cores of the most important Werner and
Lyman lines is converted to radiation at other wavelengths via
fluorescence \markcite{2042} (Black \& van Dishoeck 1987), leading to a
lower value for $\beta_{\ref{c17}}$.  The reduction in
$\beta_{\ref{c17}}$ caused by self shielding of material within
Component~1 is small: for $\log N({\rm H}_2)=14.52$ and $b=3~{\rm
km~s}^{-1}$ it is only 32\% \markcite{349} (Draine \& Bertoldi 1996). 
We will discuss in \S\ref{reconciliation} how much the photodissociation
of H$_2$ could be increased by the gas's proximity to the hot, bright
stars in the Orion association.  

\subsubsection{Expected Amount of H$_2$}\label{expected_h2}

We now investigate whether or not it is plausible that the above
reactions can produce the approximate order of magnitude of H$_2$ that
we observe in the higher $J$ levels of Component~1.  For the condition
that the preshock density $n_0 = 0.1~{\rm cm}^{-1}$ (\S\ref{shock}), we
expect that the time scale for perceptible changes in temperature and
ionization when $6500 > T > 2000$K is about $4\times 10^{11}{\rm s}$, a
value that is much greater than the equilibration time scales
$\beta_{\ref{c17}}^{-1}$ for the production of H$_2$,
$[\beta_{\ref{c8-1}} + C_{\ref{c10}}n({\rm H}^+) + C_{\ref{c9}}n({\rm
H})]^{-1}$ for H$^-$, or $[C_{\ref{c5}}n(e) + C_{\ref{c3}}n({\rm
H})]^{-1}$ for H$_2^+$.  Thus, for a total density $n_{\rm H} \equiv
n({\rm H}^+) + n({\rm H})$ and a fractional ionization $f=n({\rm
H}^+)/n_{\rm H}$ the density of H$_2$ at any particular location is
given by a straightforward equilibrium equation
\begin{mathletters}
\begin{equation}\label{h2_equilibrium}
n({\rm H}_2)=f(1-f)^2n_{\rm H}^2[F({\rm H}^-)+F({\rm H}_2^+)+F({\rm
grain})]/\beta_{\ref{c17}}
\end{equation}
with
\begin{equation}\label{FH-}
F({\rm H}^-)={C_{\ref{c9}}C_{\ref{c8}}\over \beta_{\ref{c8-1}}/n_{\rm H}
+ C_{\ref{c10}}f + C_{\ref{c9}}(1-f)}~,
\end{equation}
\begin{equation}\label{FH2+}
F({\rm H}_2^+)={C_{\ref{c3}}C_{\ref{c2}}\over C_{\ref{c5}}f +
C_{\ref{c3}}(1-f)}~,
\end{equation}
and
\begin{equation}\label{Fgrain}
F({\rm grain})=C_{\ref{cg}}/f
\end{equation}
\end{mathletters}

In order to make an initial estimate for the amount of H$_2$ that could
arise from the warm, partly ionized gas, we must evaluate the integral
of the right-hand side of Eq.~\ref{h2_equilibrium} through the relevant
part of the cooling, postshock flow.  The structure of this region is
dependent on several parameters that are poorly known and whose effects
will be discussed in \S\ref{reconciliation}.  As a starting point,
however, we can define a template for the behavior of $f$, $n_{\rm H}$
and $T$ with distance by adopting the information displayed by Shull \&
McKee \markcite{1800} (1979) for a 100~${\rm km~s}^{-1}$ shock with
$n_0=10~{\rm cm}^{-3}$ and solar abundances for the heavy elements
(their Model E displayed in Fig.~3).  To convert to our assumed
$n_0=0.1~{\rm cm}^{-3}$, we scale their densities $n({\rm H})$ and
$n({\rm H}^+)$ down by a factor of 100 and the distance scale up by the
same factor.

Over all temperatures, we discover that $F({\rm H}_2^+)$ is at least 100
times smaller than $F({\rm H}^-)$, and hence this term is not
significant for our result.  $F({\rm grain})$ is negligible compared to
$F({\rm H}^-)$ at high temperatures, but its importance increases as the
temperatures decrease: the two terms equal each other at $T=2000$K, and
$F({\rm grain})=3.7F({\rm H}^-)$ at 1000K.  Within the $F({\rm H}^-)$
term, the terms for photodestruction and recombination with H$^+$ in the
denominator are about equal at $T=6500$K, but the photodestruction
becomes much more important at lower temperatures.

The integral of the predicted $n({\rm H}_2)$ (Eq.~\ref{h2_equilibrium})
over a path that extends down to $T=2000$K equals $3.6\times 10^{13}{\rm
cm}^{-2}$.  This value is substantially lower than the amount of H$_2$
that we observed in the higher $J$ levels in Component~1
[$\sum_{J=2}^5N({\rm H}_2)=2\times 10^{14}{\rm cm}^{-2}$; see
Table~\ref{comp_summary}].

\subsubsection{Ways to Reconcile the Expected and Observed
H$_2$}\label{reconciliation}

There are several effects that can cause significant deviations from the
simple prediction for $N({\rm H}_2)$ given above.  First, if we accept
the principle that the origin of the preshock flow at $-94~{\rm
km~s}^{-1}$ is from either a stellar wind produced by $\zeta$~Ori (plus
perhaps other stars in the association) or some explosive event in
Orion, we must then acknowledge that the H$_2$ production zone is
probably not very distant from this group of stars that produce a very
strong uv flux.  As a consequence, we must anticipate that
$\beta_{\ref{c17}}$ could be increased far above that for the general
interstellar medium given in Eq.~\ref{c17}.  Eq.~\ref{h2_equilibrium}
shows that this will give a reduction in the expected yield of H$_2$ in
direct proportion to this increase.  (For a given enhancement of
$\beta_{\ref{c17}}$, we expect that the increase in $\beta_{\ref{c8-1}}$
will be very much less because the cross section for this process is
primarily in the visible part of the spectrum where the contrast above
the general background is relatively small.)  Working in the opposite
direction, however, is the fact that the stars' Lyman limit fluxes will
supplement the ionizing radiation produced by the hot part of the shock
front, thus providing heating and photoionization rates above those
given in the model.  The resulting higher level of $f$ and the increase
of the length of the warm gas zone will result in an increase in the
expected $N({\rm H}_2)$.

To see how important these effects might be, we can make some crude
estimates for the relevant increases in the uv fluxes.  In the vicinity
of 1000\AA\, i.e., the spectral region containing the most important
transitions that ultimately result in photodissociation of H$_2$, the
fluxes from $\epsilon$ and $\sigma$~Ori at the Earth are $5.6\times
10^{-8}$ and $1.8\times 10^{-8}{\rm erg~cm}^{-2}{\rm s}^{-1}{\rm
\AA}^{-1}$, respectively \markcite{1043} (Holberg et al. 1982).  We can
assume that other very luminous stars that might make important
contributions, such as $\delta$, $\zeta$, $\kappa$ and $\iota$~Ori, have
uv fluxes consistent with that of $\epsilon$~Ori after a scaling
according to the differences in visual magnitudes.  The probable
distance of the H$_2$ from the stars is probably somewhere in the range
60 to 140 pc, as indicated by various measures of the transverse
dimensions of shell-like structures seen around the Orion association
\markcite{2674} (Goudis 1982) (and assuming that the Orion association
is at a distance of 450 pc from us).  If we compare the far-uv
extinction differences for $\delta$ and $\epsilon$~Ori reported by
Jenkins, Savage \& Spitzer \markcite{1063} (1986) to these stars' color
excesses E(B$-$V) = 0.075, we infer from the uv extinction formulae of
Cardelli, Clayton \& Mathis \markcite{3280} (1989) that $R_{\rm V}$=4.6
and, again using their formulae, that $A_{\rm 1000\AA}=0.96~{\rm Mag.}$ 
In the absence of such extinction, these plus the other stars should
produce a net flux $F_{\rm 1000\AA}=1.0\times 10^{-5}r_{100}^{-2}~{\rm
erg~cm}^{-2}{\rm s}^{-1}{\rm \AA}^{-1}$, where $r_{100}$ is the distance
away from the stars divided by 100 pc.  With $r_{100}=1$,
$\beta_{\ref{c17}}$ is enhanced over the value in Eq.~\ref{c17} by a
factor of 7.

Stars in the Orion association produce about $3.8\times 10^{49}$ Lyman
limit photons ${\rm s}^{-1}$, and only a small fraction of this flux is
consumed by the ionization of hydrogen in the immediate vicinity of the
stars \markcite{3279} (Reynolds \& Ogden 1979).  From this estimate, one
may conclude that the ionizing flux of $\sim 10^6{\rm
photons~cm}^{-2}{\rm s}^{-1}$ radiated by the immediate postshock gas
\markcite{1800} (Shull \& McKee 1979) could be enhanced by a factor
approaching $30r_{100}^{-2}$, thus increasing the thickness of the
region over which there is a significant degree of ionization and
heating.

Another parameter that can influence the length of the zone where
reactions \ref{c9} and \ref{c3} are important is the relative abundances
of heavy elements.  Here, the cooling is almost entirely from the
radiation of energy by forbidden, semi-forbidden and fine-structure
lines from metals -- see Fig.~2 of Shull \& McKee \markcite{1800}
(1979).  If these elements are depleted below the solar abundance ratio
because of grain formation, the length of the warm H$_2$ production zone
must increase \markcite{3319} (Shull \& Draine 1987).  It is unlikely
that the grains will been completely destroyed as they passed through a
$90~{\rm km~s}^{-1}$ shock \markcite{2783} (Jones et al. 1994).

Finally, it is important to realize that the outcome for $N({\rm H}_2)$
should scale roughly in proportion to $n_0^2$.  The reason for this is
that over most of the path, we found that $\beta_{\ref{c8}}/n_{\rm H}$
was the most important term within denominator of the dominant
production factor $F({\rm H}^-)$.  This in turn makes $n({\rm H}_2)$
scale in proportion to $n_{\rm H}^3$ almost everywhere (note that $\int
n_{\rm H}dl$ does not vary with $n_0$).

\subsubsection{Independent Information from an Observation of
Si~II$^*$}\label{siII}

It is important to look for other absorption line data that can help to
narrow the uncertainties in the key parameters discussed above.  One
such indicator is the column density of ionized silicon in an excited
fine-structure level of its ground electronic state (denoted as
Si~II$^*$).  This excited level is populated by collisions with
electrons, and the balance of this excitation with the level's radiative
decay (and collisional de-excitations) results in a fractional abundance
\begin{equation}\label{siII*ratio}
\log\left( {N({\rm Si~II}^*)\over N({\rm Si~II})}\right) = \log n(e) -
0.5\log T_3 - 2.54
\end{equation}
\markcite{2389} (Keenan et al. 1985).  In conditions where the hydrogen
is only partially ionized, we expect that $n({\rm Si}^{++})/n{(\rm
Si}^+)$ will be much less than $n({\rm H}^+)/n({\rm H})$ because ionized
Si has a larger recombination coefficient \markcite{111} (Aldrovandi \&
P\'equignot 1973) and a smaller photoionization cross section
\markcite{1874} (Reilman \& Manson 1979) (its ionization potential of
16.34~eV is also greater than that of hydrogen).  Thus, for situations
where $f$ is not very near 1.0, it is reasonably safe to assume that
virtually all of the Si is singly ionized.  If, for the moment, we also
assume that the Si to H abundance ratio is equal to the solar value, we
expect that
\begin{equation}\label{siII*abund}
n({\rm Si~II}^*)=10^{-7}fn_{\rm H}^2T_3^{-1/2}~{\rm cm}^{-3}~.
\end{equation}
As we did for H$_2$, we can integrate the expression for $n({\rm
Si~II}^*)$ through the modeled cooling zone to find an expectation for
the column density $N({\rm Si~II}^*)=5.8\times 10^{11}{\rm cm}^{-2}$.  

A very weak absorption feature caused by Si~II$^*$ at approximately the
same velocity as our Component~1 can be seen in a medium resolution HST
spectrum\footnote{Archive exposure identification Z165030GT} of
$\zeta$~Ori~A that covers the very strong transition at 1264.730\AA. 
Our measurement of this line's equivalent width was $13.4\pm 3$m\AA,
leading to $N({\rm Si~II}^*)=1.0\times 10^{12}{\rm cm}^{-2}$, a
result\footnote{From the equivalent width of 3.4m\AA\ (no error stated)
for the 1194.49\AA\ line of Si~II$^*$ reported by Drake \& Pottasch
\markcite{1803} (1977), one obtains a somewhat lower value, $\log N({\rm
Si~II}^*)=11.6$} that is almost twice the prediction stated above.

From our result for Si~II$^*$, we conclude that the combined effect of
the stars' ionizing flux and a possible increase in $n_0$ over our
assumed value of $0.1~{\rm cm}^{-3}$ could raise $\int n(e)n_{\rm H}dl$
by not much more than a factor of two.  However, we have no sensitivity
to the possibility that metals are depleted since the decrease in the
abundance of Si would be approximately compensated by the increase in
the characteristic length for the zone to cool (assuming the primary
coolants and Si are depleted by about the same amount).  Thus, it is
still possible that the our calculation based on a model with solar
abundances will result in an inappropriate (i.e., too low) value for the
expected $N({\rm H}_2)$. 

\subsubsection{Coupling of the Rotational Temperature to
Collisions}\label{coupling}

One remaining task is to establish that the conditions in the
H$_2$-formation zone are such that collisional excitation of the higher
$J$ levels can overcome the tendency for the molecules to move to other
states through either radiative decay or the absorption of uv photons. 
Tawara  et al. \markcite{3266} (1990) summarize the collision cross
sections as a function of energy for excitations $J=0\rightarrow 2$ and
$J=1\rightarrow 3$ by electrons.  We calculate that these cross sections
should give a rate constant of about $1\times 10^{-10}T_3^{3/2}{\rm
cm}^3{\rm s}^{-1}$ for $T_3\gtrsim 2$.  Thus, for $n_e\gtrsim 1~{\rm
cm}^{-3}$ and $T_3\approx 5$ the collisions can dominate over radiative
transition rates of about $3\times 10^{-10}~{\rm s}^{-1}$ for $J=2$ and
3.  To collisionally populate $J=5$ which can decay at a rate of
$1\times 10^{-8}~{\rm s}^{-1}$ to $J=3$, we would need to have
$n_e\gtrsim 10~{\rm cm}^{-3}$ just to match the radiative rate, assuming
that the collisional rate constant is not significantly lower than what
we calculated for $J=2$ and 3.

\subsection{Further Formation of H$_2$ in a Cool
Zone}\label{cool_formation}

Additional formation of H$_2$ molecules probably takes place in gas that
has cooled well below 2000K and is nearly fully recombined.  We are
unable to distinguish between this gas and the material that was
originally present as an obstruction to the high velocity flow to create
the bow shock.  To obtain an approximate measure of the total amount of
cool, mostly neutral gas, we determined $\int N_a(v)dv$ over the
velocity interval $-10 < v < +5~{\rm km~s}^{-1}$ for the N~I line at
1134.165\AA\ which does not appear to be very strongly saturated.  The
relative ionization of nitrogen should be close to that of hydrogen
\markcite{1927} (Butler \& Dalgarno 1979), and this element is not
strongly depleted in the interstellar medium \markcite{14} (Hibbert,
Dufton, \& Keenan 1985).  Our conclusion that $\log N({\rm N~I})=14.78$
leads to an inferred value for $\log N({\rm H})$ equal to 18.73. 
According to a model\footnote{This model is not exactly applicable to
our situation, since it has a compression ratio of 4 instead of our
value of 2.6} for a 90~km~s$^{-1}$ shock of Shull \& McKee
\markcite{1800} (1979), $\log N({\rm H})=18.40$ is the amount of
material that accumulates by the time $T$ reaches 1000K.  Hence, from
our measure of the total N(H) (but indeed an approximate one), we
estimate that the amount of gas at $T<1000$K is about comparable to that
at the higher temperatures. 

An insight on the conditions in the cool, neutral zone is provided by
the populations of excited fine-structure levels of C~I.  Jenkins \&
Shaya \markcite{1034} (1979) found that $\log p/k=4.1$ in the part of
our Component~1 that carries most of the neutral carbon atoms.  If we
take as a representative temperature $T=300$K, the local density should
be $n_{\rm H}\approx 40~{\rm cm}^{-3}$ and $C_{\ref{cg}}=1.8\times
10^{-17}{\rm cm}^3{\rm s}^{-1}$.  With the general interstellar value
for $\beta_{\ref{c17}}$, we expect an equilibrium concentration $n({\rm
H}_2)/n_{\rm H}=2\times 10^{-5}$.  When we multiply this number by our
estimate $N({\rm H})=3\times 10^{18}{\rm cm}^{-2}$, we find that we
should expect to observe $\log N({\rm H}_2)=13.8$, a value that, after
considering the crudeness of our calculations, is acceptably close to
our actual measurements of H$_2$ in $J=0$ and 1, the levels that arise
primarily from the coolest gas.  If $\beta_{\ref{c17}}$ is significantly
enhanced by radiation from the Orion stars (\S\ref{reconciliation}), we
would then have difficulty explaining the observations.

\section{Summary}\label{summary}

We have observed over 50 absorption features in the Lyman and Werner
bands of H$_2$ in the uv spectrum of $\zeta$~Ori~A.  An important aspect
of our spectrum is that it had sufficient resolution to detect in one of
the velocity components (our Component~1) some important changes in the
one-dimensional velocity distributions of the molecules with changing
rotational excitation $J$.  The main focus of our investigation has been
to find an explanation for this result, since it is a departure from the
usual expectation that the rotational excitation comes from uv pumping,
an effect that would make the profiles look identical.  A smaller amount
of broadening for higher $J$ is also seen for Component~3, a component
that has much more H$_2$ than Component~1.

In Component~1, we have found that as $J$ increases from 0 to 5 there is
a steady increase in the width of the velocity profile, combined with a
small drift of the profile's center toward more negative velocities.  We
have shown that the pumping lines are not strong enough to make
differential shielding in the line cores a satisfactory explanation for
the apparent broadening of the $J$ levels that are populated by such
pumping.  While one might resort to an explanation that unseen,
additional H$_2$ could be shielding light from a uv source (or sources)
other than $\zeta$~Ori, we feel that this interpretation is implausible,
and, moreover, it does not adequately explain the differences that we
see between the profiles of $J=0$ and 1.

One could always argue that the absorption that we identify as
Component~1 is really a chance superposition of two, physically
unrelated regions that have different rotation temperatures and central
velocities.\footnote{There is a good way to illustrate how this creates
the effect that we see in Component~1.  Imagine that we recorded the
H$_2$ lines at a resolution that was so low that Complexes 1 and 3 were
not quite resolved from each other.  We would then see features that got
broader with increasing $J$, and their centers would shift toward the
left.  This is qualitatively exactly the same effect that we see on a
much smaller velocity scale within Component~1.}  While not impossible,
this interpretation is unattractive.  It requires a nearly exact
coincidence of the two regions' velocity centers to make up a component
that stands out from the rest of the H$_2$ absorption and, at the same
time, shows smoothly changing properties with $J$ at our velocity
resolution.  We feel that the most acceptable interpretation is the
existence of a coherent region of gas that, for some particular reason
that has a rational explanation, has systematic changes in the
properties of the material within it that could produce the effects that
we observe.  One phenomenon that fits this picture is the organized
change in temperature and velocity for gas that is cooling and
recombining in the flow behind a shock front.  The excess width of the
higher $J$ lines could arise from both higher kinetic temperatures and
some velocity shear caused by the steady compression of the gas as it
cools.

Our concept of a shock is supported by evidence from atomic absorption
lines in the spectrum of $\zeta$~Ori~A.  We see features that we can
identify with both the preshock medium and the immediate postshock gas
that is very hot.  If this interpretation is correct (and not a
misguided attempt to assign a significant relationship between atomic
components with different levels of ionization at very different
velocities), we can use the atomic features to learn much more about the
shock's general properties.

We start with the expectation that the coolest molecular material, that
which shows up in the lowest $J$ levels at $v=-1~{\rm km~s}^{-1}$, is in
a region containing gas that is very strongly compressed and thus nearly
at rest with respect to the shock front.  The atomic features of C~II
and N~II that we identify with the preshock flow appear at a velocity of
$-94~{\rm km~s}^{-1}$ with respect to the cool H$_2$.  Hence this is the
value that we normally associate with the ``shock velocity.'' This
preshock gas also shows up in the lines of C~III, N~III, Si~III, S~III
and Al~III.  An upper limit to its temperature $T<20,000$K results from
the apparent lack of C~IV that would arise from collisional ionization
at slightly higher temperatures.  The temperature could be as low as
typical H~II region temperatures ($8,000<T<12,000$K) if uv photons are
the main source of ionization.

Absorption features from more highly ionized gas at around $-36~{\rm
km~s}^{-1}$ that show up in the C~IV doublet indicate that the initial
compression factor is only 2.6, a value that is significantly lower than
the usual 4.0 expected for a shock with a high Mach number.  Reasonable
numbers for the preshock density, temperature and magnetic field
strength ($0.1~{\rm cm}^{-3}$, 20,000K and $10\mu$G) can explain this
lower compression factor and establish a switch-on shock.  However,
except for a limit $n_0<0.3~{\rm cm}^{-3}$ that comes from the lack of
C~II$^*$ absorption, we have no independent information that can
distinguish between these somewhat arbitrary assignments and other,
equally acceptable combinations.  

To overcome the problem that there seems to be a low initial compression
of the gas but eventually the densities are allowed to increase to the
point that H$_2$ can form, we invoke the concept of an oblique magnetic
shock, where the theoretical models outline the existence of two
discontinuities, a ``switch on'' front and a ``switch off'' front. 
However, we do not attempt to explore the validity of this picture in
any detail.

Neglecting complications that are introduced by the oblique shock
picture, we expect that as the gas flow cools to temperatures
significantly below the immediate postshock temperature, it decelerates
and begins to show ions that have ionization potentials below that of
C~IV, such as Si~III (at $-20~{\rm km~s}^{-1}$) and Al~III ($-15~{\rm
km~s}^{-1}$).  At temperatures somewhat below $10^4$K, the gas should be
still partially ionized and at a density $n_{\rm H}>60n_0$.  These
conditions favor the production of H$_2$ via the formation of H$^-$ and
its subsequent reaction with H to produce the molecule plus an electron,
rather than the usual formation on grains that dominates in cool clouds. 
Our observation of the Si~II$^*$ absorption feature at a velocity near
$0~{\rm km~s}^{-1}$ indicates that it is unlikely that $n_0\ll 0.1~{\rm
cm}^{-3}$, and thus the density in the molecule forming region is high
enough to insure that the photodetachment of H$^-$ does not deplete this
feed material to the point that the expected abundance of H$_2$ is well
below the amount that we observe.

It is possible that the uv flux from the Orion stars could enhance the
H$_2$ photodissociation rate $\beta_{\ref{c17}}$ to a level that is far
above that which applies to the average level in our part of the Galaxy. 
If this is true, then we must make a downward revision to our prediction
that $\log N({\rm H}_2)=13.56$. At the same time, however, additional
ionizing photons from the stars could lengthen the warm H$_2$ production
zone, and this effect may gain back a large amount of the lost H$_2$.  A
reduction in the metal abundance in the gas may also lengthen the zone,
giving a further increase in the expected H$_2$.

As the gas compresses further and becomes almost fully neutral, the
H$^-$ production must yield to grain surface reactions as the most
important source of molecules.  Using information from the C~I
fine-structure excitation, we can infer that the density in the cool gas
is sufficient to give $n({\rm H}_2)/n_{\rm H}$ equal to about half of
what we observed, if we assume that most of the H$_2$ in the $J=0$ and 1
states comes from the cool region.

On the basis of a diverse collection of evidence and some rough
quantitative calculations, we have synthesized a general description of
the cooling gas behind the shock and have shown that H$_2$ production
within it could plausibly explain the unusual behavior in the profiles
that we observed.  Obviously, if one had the benefit of detailed shock
models that incorporated the relevant magnetohydrodynamic, atomic and
molecular physics, it would be possible to substantiate this picture (or
perhaps uncover some inconsistencies?) and narrow the uncertainties in
various key parameters.  Also, more detailed models should allow one to
address certain questions that are more difficult to answer, such as
whether or not more complex chemical reactions play an important role in
modifying the production of H$_2$; we have identified only a few good
prospects.  For instance, is there enough Ly-$\alpha$ radiation produced
in the front (or in the H~II region ahead of it) to make the formation
by excited atom radiative association (i.e., H($n=1$) + H($n=2$)
$\rightarrow$ H$_2$ + $h\nu$) an important additional production route
\markcite{1923} (Latter \& Black 1991)?  On the observational side, we
expect to see very soon a vast improvement in the amount and quality of
data on atomic absorption lines toward $\zeta$~Ori~A.  Very recently,
the GHRS echelle spectrograph on HST obtained observations of various
atomic lines at extraordinarily good resolution and S/N.

\acknowledgments

Support for flying IMAPS on the ORFEUS-SPAS-I mission and the research
reported here came from NASA Grant NAG5-616 to Princeton University. 
The ORFEUS-SPAS project was a joint undertaking of the US and German
space agencies, NASA and DARA.  The successful execution of our
observations was the product of efforts over many years by engineering
teams at Princeton University Observatory, Ball Aerospace Systems Group
(the industrial subcontractor for the IMAPS instrument) and Daimler-Benz
Aerospace (the German firm that built the ASTRO-SPAS spacecraft and
conducted mission operations). Most of the development of the data
reduction software was done by EBJ shortly after the mission, while he
was supported by a research award for senior U.S. scientists from the
Alexander von Humboldt Foundation and was a guest at the Institut f\"ur
Astronomie und Astrophysik in T\"ubingen. We are grateful to B.~T.
Draine for valuable advice about the different alternatives for
interstellar shocks.  B.~T. Draine, L. Spitzer, and J.~H. Black supplied
useful comments on an early draft of this paper.  Some of the
conclusions about atomic absorption features are based on observations
made with the NASA/ESA Hubble Space Telescope, obtained from the data
archive at the Space Telescope Science Institute.  STScI is operated by
AURA under NASA contract NAS 5-26555.

\newpage
\end{document}